\documentclass[preprint]{elsarticle}

\usepackage{graphicx}
\usepackage{amsmath,amssymb}
\usepackage{bm}
\usepackage{ascmac}
\usepackage{dcolumn}
\usepackage{multicol}	 
\usepackage{theorem}
\usepackage{wrapfig}
 \usepackage{mathtools}
 \usepackage{xcolor}
\usepackage{txfonts}

\mathtoolsset{showonlyrefs}
\mathtoolsset{centercolon}

\newcommand{\R}{\varmathbb{R}}

\newcommand{\inta}{\int_{\R^3} \int_0^{+\infty}}
\newcommand{\dA}{\, \varphi(\mathcal{I}) \, d\mathcal{I} d {\boldsymbol C}}
\newcommand{\da}{\, \varphi(\mathcal{I}) \, d\mathcal{I} d {\boldsymbol \xi}}
\newcommand{\cc}{{\boldsymbol \xi}}
\newcommand{\CC}{{\boldsymbol C}}
\newcommand{\xx}{\mathbf{x}}

\newcommand{\PP}{\mathcal{P}}
\newcommand{\II}{\mathcal{I}}
\newcommand{\bei}{{\bar{\varepsilon}^I}}
\newcommand{\cvi}{{\hat{c}_v^I}}
\newcommand{\EE}{{\textemdash E}}

\newcommand{\difg}[2]{\ifx#2\rho{#1_{\rho}}\else{#1_{\hat{G}_{ll}}}\fi}
\newcommand{\dift}[2]{\ifx#2\rho{\left(\frac{\partial #1}{\partial \rho}\right)_T}\else{\left(\frac{\partial #1}{\partial T}\right)_\rho}\fi}

\journal{International Journal of Non-Linear Mechanics }

\begin{document}

\begin{frontmatter}

\title{Which  moments are appropriate to describe  gases with internal structure  in Rational Extended Thermodynamics?}

\author{Takashi Arima}
\ead{arima@tomakomai-ct.ac.jp}
\address{ Department of Engineering for Innovation National Institute of Technology, Tomakomai College, Tomakomai, Japan}

\author{Maria Cristina Carrisi}
\ead{mariacri.carrisi@unica.it}
\address{Department of Mathematics and Informatics, University of Cagliari, Cagliari, Italy}

\author{Sebastiano Pennisi}
\ead{spennisi@unica.it}
\address{Department of Mathematics and Informatics, University of Cagliari, Cagliari, Italy}

\author{Tommaso Ruggeri\corref{mycorrespondingauthor}}
\cortext[mycorrespondingauthor]{Corresponding author}
\ead{tommaso.ruggeri@unibo.it}
\address{Department of Mathematics and AM$^2$, University of Bologna, Italy\\
\vspace{1cm} \hspace{6cm} Dedicated to Luigi Preziosi for his 60 birthday}

\begin{abstract}
Motivated by a recent paper of Pennisi in the relativistic framework \cite{Pennisi-2021},  we revisit the previous approach of two hierarchies of moments critically and propose a new natural physical hierarchy of moments to describe classical rarefied non-polytropic polyatomic gas in the framework of  Molecular Rational Extended Thermodynamics. The differential system of the previous approach is proved to be a  \emph{principal subsystem} of the present one. The main idea is that at the molecular level, the total energy is the kinetic energy plus the energy of internal mode due to the rotation and vibration, and the increasing moments contain this total energy as power in which the power index increases with the number of tensorial indexes. In particular, we consider the case of $15$ moments, and we close the system using the variational method of the Maximum Entropy Principle. We prove the convexity of entropy and the possibility to put the system in symmetric form.  This more rich kinetic framework may be interesting also as possible applications to biomathematics or other fields in which kinetic models were applied recently with success.
\end{abstract}

\begin{keyword}
Rational extended thermodynamics\sep Non-equilibrium thermodynamics\sep  Rarefied polyatomic gases
\MSC[2020] 35L40 \sep  76N10 \sep 76N15
\end{keyword}

\end{frontmatter}

\section{Introduction}
The kinetic theory offers an excellent  model for highly rarefied gases. The celebrated Boltzmann equation is widely used in many applications and is still now a challenge for its difficult mathematical questions. Cercignani was one of the world leaders that gave fundamental papers on this subject that are collected in the books \cite{Cer1,Cer2}.
More recently, the Kinetic theory was used in fields very far from gas dynamics like in biological phenomena, socio-economic systems, models of swarming, and many other fields (see, for example, \cite{Preziosi1,Tosin,Carillo} and references therein).

Rational Extended Thermodynamics (RET) is a theory that wants to offer a phenomenological model that is a sort of bridge between the Navier-Stokes-Fourier theory and the Boltzmann equation. RET is strictly related to the molecular approach, so-called molecular ET which adopts the Maximum Entropy Principle (MEP) as the closure of moments associated with the distribution function. This theory is described in the two editions of the book of M\"uller and Ruggeri  \cite{ET,RET} and is substantially limited to the monatomic gas as the original Boltzmann equation.

More recently, an extension of RET was given to include polyatomic and dense gases. This new approach starts with the paper of Arima, Taniguchi, Ruggeri, and Sugiyama \cite{Arima-2011} and is strictly related to the extension of the kinetic theory of polyatomic gas thanks to the refreshed idea given by Borgnakke and Larsen \cite{Borgnakke-1975}  and the mathematical treatment due to the French mathematicians Bourgat, Desvillettes, Le Tallec, and Perthame \cite{Bourgat-1994}. The state of the art on this subject is summarized in the two books of Ruggeri and Sugiyama \cite{book,newbook}.

This paper aims to give first a brief historical summary of this new approach for rarefied gases and to recognize in the case of many moments some critical limitations on the choice of the moments. Starting from a new idea of Pennisi in the relativistic framework \cite{Pennisi-2021}, we will propose a new more physical choice of moments and we discuss in particular the simplest case of $15$ fields as an example of the new approach. Such a more sophisticated model including also internal mode such as rotation and vibration of a molecule might be useful to use the gas theory as an analogy for other scientific branches as mentioned before.

\section{A brief  survey of moment hierarchies in Rational Extended Thermodynamics }
To understand the aim of the present paper, it  is necessary to give a brief survey of the classical and relativistic structure of RET for rarefied gases associated with the moments of kinetic theory. The interested readers can find more details in the recent book of Ruggeri and Sugiyama \cite{newbook}. 
\subsection{Monatomic Gas}
In the phenomenological Rational Extended Thermodynamics (RET) \cite{RET}  of  monatomic gas, there exists a single hierarchy of field equations:
\begin{equation}\label{momentini}
\frac{\partial F^M_{k_1 k_2 \dots k_n}}{\partial t}+ \frac{\partial F^M_{k_1 k_2 \dots k_n k_{n+1}}}{\partial x_{k_{n+1}}} = P^M_{k_1k_2 \dots k_n}, \qquad n=0,1,\dots , \bar{N},
\end{equation} 
that was motivated by the moments structure associated to the Boltzmann equation
\begin{equation} \label{eq:Boltzmann}
 	\frac{\partial f}{\partial t} + \xi_i\, \frac{\partial  f}{\partial x_i} = Q(f),
 \end{equation}
 in which 
  \begin{equation*}
  F^M_{k_1 k_2 \dots k_n}(\xx,t) = m \int_{\R^3} f(\xx,t,\bm{\xi})  \, \xi_{k_1}  \xi_{k_2} \dots  \xi_{k_n} \, d\cc , \qquad k_1,k_2, \dots, k_n= 1,2,3  ,
  \end{equation*}
  and 
  \begin{equation*}
  P^M_{k_1k_2 \dots k_n} = m \int_{\R^3} Q(f)\, \xi_{k_1}  \xi_{k_2} \dots  \xi_{k_n}  \, d\cc, \qquad P^M=P^M_{k_1}=P^M_{kk}=0,
  \end{equation*}
 where the state of the gas can be described by the distribution function $f(\xx, t, \cc)$, 
being respectively $\xx \equiv(x_i),\cc\equiv(\xi_i),t$ the space coordinates, the microscopic velocity and the time.  $Q$ denotes the collisional term, 
 $m$ is the atomic mass, and the moment with $n=0$, denoted as $F^M$, is the mass density $\rho$.

 To obtain the closed set of field equations for the moment system  \eqref{momentini}, it is necessary to find the constitutive theory for the last flux and production terms. For example, the Euler system, whose fields are the mass density $F^M(=\rho)$, momentum density $F^M_i=\rho v_i$ where $v_i$ is the velocity, and energy density $F^M_{ll}=\rho v^2 + 2\rho \varepsilon $ where $\varepsilon$ is the internal energy and $v^2= v_i v^i= v_1^2+v_2^2+v_3^2$, is the case with $\bar{N}=2$ taking the trace of the second-order tensor. The case $\bar{N}=3$ taking the trace of the third-order tensor (ET$_{13}$) corresponds to the well-known  Grad 13-field theory \cite{Grad} whose fields are the traceless part of momentum flux\footnote{The angle brackets indicate the traceless part of the tensor $U_{ij}$: $U_{\langle ij\rangle} = U_{ij} - \frac{1}{3}U_{ll}\delta_{ij}$.}  $F^M_{\langle ij\rangle}$and energy flux $F^M_{lli}$ in addition to $F^M, F^M_i$ and $F^M_{ll}$. 
 
 \bigskip
 
 This situation has a relativistic counterpart. The first relativistic version of the modern RET was given by Liu, M\"uller and Ruggeri (LMR) \cite{LMR} considering the Boltzmann-Chernikov relativistic equation
  \cite{BGK,Synge,KC}:
  \begin{equation}\label{BoltzR}
  p^\alpha \partial_\alpha f = Q,
  \end{equation}
 in which the distribution function  $f$ depends on  $(x^\alpha,  p^\beta)$, where $x^\alpha$ are the space-time coordinates, $p^\alpha$ is the four-momentum, $\partial_{\alpha} = \partial/\partial x^\alpha$, $c$ denotes the light velocity, $m$ is the particle mass in the rest frame and $\alpha, \beta =0,1,2,3$. 
 The relativistic moment equations associated with \eqref{BoltzR}, truncated at tensorial index $N+1$, are now\footnote{When $n=0$, the tensor reduces to $A^\alpha$. Moreover, the production tensor in the right-side of \eqref{RelmomentMono} is zero for $n=0,1$,  because the first $5$ equations   represent the conservation laws of the particle number and the energy-momentum, respectively.}:
  \begin{equation}\label{Relmomentseq}
  \partial_\alpha A^{\alpha \alpha_1 \cdots \alpha_n  } =  I^{  \alpha_1 \cdots \alpha_n   }
  \quad \mbox{with} \quad n=0 \, , \,\cdots \, , \,  N 
  \end{equation}
  with  
  \begin{align}\label{RelmomentMono}
  A^{\alpha \alpha_1 \cdots \alpha_n  } = \frac{c}{m^{n-1}} \int_{\R^{3}}
  f  \,  p^\alpha p^{\alpha_1} \cdots p^{\alpha_n}  \, \, d \boldsymbol{P}, \quad I^{\alpha_1 \cdots \alpha_n  } = \frac{c}{m^{n-1}} \int_{\R^{3}}
  Q  \,   p^{\alpha_1} \cdots p^{\alpha_n}  \, \, d \boldsymbol{P} ,
  \end{align}
 and 
  \begin{equation*}
  d \boldsymbol{P} =  \frac{dp^1 \, dp^2 \,
  	dp^3}{p^0} .
  \end{equation*}
  When $N=1$, we have the relativistic Euler system, and when $N=2$, we have the LMR theory of a relativistic gas with $14$ fields:
  \begin{equation}
   \partial_\alpha A^{\alpha } = 0, \quad \partial_\alpha A^{\alpha \beta} =0, \quad  \partial_\alpha A^{\alpha \beta \gamma} =  I^{  \beta \gamma  }, \qquad \left(\beta,\gamma=0,1,2,3; \,\, I^\alpha_{\,\,\alpha} =0 \right). \label{Annals}
   \end{equation}
   In the relativistic case, differently from the classic one, we need to take all indexes in \eqref{Relmomentseq} because when we take the trace pair of two or more indexes in the moments of \eqref{RelmomentMono}, we obtain a lower order tensor due to the constraint on the four-momentum: $p_\alpha p^\alpha = m^2 c^2$.
   
   In a recent paper, Pennisi and Ruggeri \cite{PRS} (see also \cite{newbook}) proved that the classical limit of \eqref{Relmomentseq} gives a precise single hierarchy of the moment equations of classical case \eqref{momentini}. In fact, they proved that, for a given $N$, there exists an integer $s=0,1,\dots,N$ such that the moments expression in the classical limit is:
   \begin{equation}\label{LMono}
   F^{s|M}_{i_1 i_2\dots i_{N-s}} \equiv F^M_{j_1 j_1\dots j_s j_s i_1 i_2\dots i_{N-s}} = m \int_{\R^3} f(\xx,t,\bm{\xi})  \, \xi^{2s} \xi_{i_1}  \xi_{i_2} \dots  \xi_{N-s} \, d\cc , \quad s=0,1,\dots,N,
   \end{equation}
where $\xi^2 = \xi_1^2 + \xi_2^2 + \xi_3^2$. This means that for $s=0$ all the $N$ indexes are different and then we have contraction of two indexes as $s$ increases until the highest moment is a scalar because all the indexes are contracted. The importance of this result is that $\bar{N} = 2 N$ and then $\bar{N}$ is even and the integrals defining the moments can be integrable.  Concerning the integrability of moments see \cite{Levermore,Boillat-1997}.  According with this general result, the LMR theory converges, in the classical limit, to Kremer's monatomic ET$_{14}$ theory \cite{Kremer14}, not Grad's theory (thus ET$_{13}$) as was proved also in previous papers \cite{Weiss-Dreyer,RET,PRS}.

%


\subsection{Polyatomic Gas}
Although RET was well-established and applied to the study of the linear and non-linear waves \cite{book}, its applicability range was limited to monatomic gases. After some previous attempts \cite{Liu,KremerPoly}, RET for rarefied polyatomic gases was proposed from the phenomenological viewpoint by Arima, Taniguchi, Ruggeri and Sugiyama \cite{Arima-2011}. Because of the dynamic pressure related to the relaxation of the molecular internal modes which was absent in monatomic gases, the number of fields is now $14$ (ET$_{14}$), which dynamics is described by a binary hierarchy of field equations:
  \begin{align} \label{ET14poly}
    & \frac{\partial F}{\partial t} +  \frac{\partial F_i}{\partial x_i} =0, \nonumber\\
    & \frac{\partial F_j}{\partial t} +  \frac{\partial F_{ij}}{\partial x_i} =0,\nonumber\\
    & \frac{\partial F_{ij}}{\partial t} +  \frac{\partial F_{ijk}}{\partial x_k} =P_{ ij},
    && \frac{\partial G_{ll}}{\partial t} +  \frac{\partial G_{llk}}{\partial x_k} = 0, \\
    & && \frac{\partial G_{lli}}{\partial t} +  \frac{\partial G_{llik}}{\partial x_k} =Q_{lli},\nonumber
  \end{align}
  where $F(=\rho)$ is the mass density, $F_i(=\rho v_i)$ is the momentum density, $G_{ll}=\rho v^2 +2 \rho\varepsilon$ is two times the energy density, $F_{ij}$ is the momentum flux, and
  $G_{llk}$ is the energy flux. Differently  from the case of monatomic gases, $\varepsilon$ includes the energy of the internal mode.   $F_{ijk}$ and $G_{llik}$ are the fluxes of $F_{ij}$ and $G_{lli}$,  respectively, and $P_{ij}$ ($P_{ll}\neq 0$) and $Q_{lli}$  are the productions with respect to $F_{ij}$ and  $G_{lli}$, respectively.
In the monatomic singular limit, it converges to the monatomic ET$_{13}$ theory \cite{Arima-2013,book,newbook}.

The binary hierarchy was justified and also derived from molecular ET \cite{Arima-2013,Pavic-2013} by using a kinetic model where the distribution function $f$ depends on an extra variable $\II$ that takes into account the internal degrees of freedom of a molecule such as rotation and vibration \cite{Borgnakke-1975,Bourgat-1994}, i.e., $f \equiv
 f(\xx,t,\cc,\II)$, where $f(\xx,t,\cc,\II)\, d\xx\, d\cc$ is the number density of molecules with the energies $\II$ at time $t$ and in the volume element $d\xx\, d\cc$ of the phase space ($6D$ position-velocity space) centered at $\left(\xx,\cc\right) \in \R^3\times \R^3$. 
The Boltzmann equation is formally the same as the one of monatomic gases \eqref{eq:Boltzmann}, but, for the collision term $Q(f)$, we take into account the influence of internal degrees of freedom through the collision cross-section.  In this case, the macroscopic quantities such as $F$, $F_i$,$F_{ij}$, $G_{ll}$ and $G_{lli}$, are now the moments defined as  follows:
\begin{align}
  \begin{split}
 &  F_{i_1 \ldots i_j} = m \inta  f \, \xi_{i_1}\cdots \xi_{i_j}  \,  \da ,    \\ 
 &   {G_{lli_1 \ldots i_k}}  =  2\inta f \, \left(\frac{m \xi^2}{2} + \II\right) \xi_{i_1}\cdots \xi_{i_k}  \,  \da .
  \end{split}
  \label{mom3}
\end{align}
Here $\varphi(\II)$ is the state density corresponding to $\II$, i.e., $\varphi(\II) d\II$ represents the number of internal state between $\II$ and $\II + d\II$.  As it can be easily seen in \eqref{mom3}, the moments $F_{i_1 \ldots i_j}$ are free from the microscopic energy $\II$ and the $G_{lli_1 \ldots i_k}$ are moments of the sum of the  microscopic kinetic energy $m\xi^2/2$ and the microscopic internal energy $\II$. As $G_{ll}$ is the energy, except for a factor $2$, we have that the $F's$ are the usual \emph{momentum-like} moments  and the $G's$ are \emph{energy-like} moments.

From the Boltzmann equation \eqref{eq:Boltzmann}, we obtain a binary hierarchy of balance equations \eqref{ET14poly}. After the derivation of the 14-field theory \cite{Pavic-2013}, the theory with the binary hierarchy was generalized to the case with any number of moments \cite{Arima-2014,Arima-2014Lincei} with the so called $(F,G)$-hierarchies:
\begin{align}\label{momentiniFG}
\begin{split}
& \frac{\partial F_{k_1 k_2 \dots k_n}}{\partial t}+ \frac{\partial F_{k_1 k_2 \dots k_n k_{n+1}}}{\partial x_{k_{n+1}}} = P_{k_1k_2 \dots k_n}, \qquad \quad \,\,\,\, n=0,1,\dots , \bar{N}, \\
& \frac{\partial G_{ll k_1 k_2 \dots k_m}}{\partial t}+ \frac{\partial G_{ll k_1 k_2 \dots k_m k_{m+1}}}{\partial x_{k_{m+1}}} = Q_{ll k_1k_2 \dots k_m}, \qquad m=0,1,\dots , \bar{M}.
\end{split}
\end{align} 
From the requirement of the Galilean invariance and the physically reasonable solutions, it is shown that $\bar{M}=\bar{N}-1$. The case with $\bar{N}=1$ corresponds to the Euler system, and the one with $\bar{N}=2$ corresponds to ET$_{14}$.

Recently, Pennisi and Ruggeri first constructed a relativistic version of polyatomic ET theory with \eqref{Relmomentseq}  in the case of $N=2$ \cite{Annals} (see also \cite{Car1,Car2}) whose moments are given by
\begin{align} 
\begin{split}
& A^\alpha \equiv V^\alpha = m c \inta f p^\alpha \phi(\mathcal{I}) \, d \mathcal{I} \, d \boldsymbol{P}  \, , \\
& A^{\alpha \beta} \equiv T^{\alpha\beta} = \frac{1}{mc} \inta f p^\alpha p^\beta (mc^2 + \II) \, \phi(\mathcal{I}) \, d \mathcal{I} \, d \boldsymbol{P}  \, , \\
& A^{\alpha \beta \gamma  } = \frac{1}{m^2 c} \int_{\R^{3}}
\int_0^{+\infty} f  \,   p^{\alpha} p^\beta p^{\gamma}  \, \left( mc^2 + 2\II \right) \, 
\phi(\mathcal{I}) \, d \mathcal{I} \, d \boldsymbol{P}  \, , 
\end{split}
\label{PS3}
\end{align}
where a distribution function $f(x^\alpha, p^\beta,\mathcal{I})$ depends on the extra energy variable $\mathcal{I}$, similar to the classical one. In \cite{Annals}, by taking $A^{\alpha \langle \beta \gamma \rangle}$ instead of $A^{\alpha\beta\gamma}$ in \eqref{PS3}$_3$ as fields, the relativistic theory with 14 fields was proposed. It was also shown that its classical limit coincides with ET$_{14}$ based on the binary hierarchy \eqref{ET14poly} \cite{Arima-2011}. The beauty of the relativistic counterpart is that there exists a single hierarchy of moments, but, as was noticed by the authors, to obtain the classical theory of ET$_{14}$, it was necessary to put the factor 2 in front of $\II$  in the last equation of \eqref{PS3}!
This was also more evident in  the theory with any number of moments where Pennisi and Ruggeri generalized \eqref{PS3} considering the following moments \cite{PRS}:
\begin{align} 
\begin{split}
& A^{\alpha \alpha_1 \cdots \alpha_n  } = \frac{1}{m^n c} \int_{\R^{3}}
\int_0^{+\infty} f  \,  p^\alpha p^{\alpha_1} \cdots p^{\alpha_n}  \, \left(mc^2 +  n \II \right)\, 
\phi(\mathcal{I}) \, d \mathcal{I} \, d \boldsymbol{P}  \, , \\
& I^{\alpha_1 \cdots \alpha_n  } = \frac{1}{m^{n}c} \int_{\R^{3}}
\int_0^{+\infty} Q  \,  p^{\alpha_1} \cdots p^{\alpha_n}  \, \left( mc^2 +  n\II \right)\, 
\phi(\mathcal{I}) \, d \mathcal{I} \, d \boldsymbol{P}. \\
\end{split}
\label{relRETold}
\end{align}
In this case, we need a factor $n \II$ in \eqref{relRETold} to obtain in the classical limit the $(F,G)$-hierarchies \eqref{momentiniFG} .
We remark that when $N=2$, the $15$ moments theory with the triple hierarchy and not the binary one is obtained in the classical limit, because now the third order tensor is $A^{\alpha   \beta \gamma}$ and not its traceless part. Based on \eqref{relRETold}, the same authors studied the classical limit in \cite{PRS} and obtained the following set of classical moments equations:
\begin{align}
 \begin{split}
 &\partial_t H^*{}^s_{i_1 \cdots i_h} + \partial_i H^*{}^s_{ii_1\cdots i_h} = J^*{}^s_{i_1\cdots i_h}  \\ \label{sysgenold}
 &\text{for } \ s=0, \cdots , N, \ \text{and} \ h=0, \cdots, N-s,
 \end{split}
\end{align}
where
\begin{align}
 \begin{split}
 & H^*{}^s_{i_1 \cdots i_h} = %
  2\displaystyle  \inta f \, \xi_{i_1}\cdots \xi_{i_h} \, \xi^{2(s-1)}\, \left(\frac{m\xi^2}{2} + s \II\right) \da,  \\
 & J^*{}^s_{i_1 \cdots i_h}  = %
  2\displaystyle  \inta Q \, \xi_{i_1}\cdots \xi_{i_h} \, \xi^{2(s-1)}\, \left(\frac{m\xi^2}{2} + s \II\right) \da . 
\\
 \end{split}
\label{Hsold}
\end{align}
The $H^*$'s hierarchy coincides with $F$'s hierarchy $\eqref{mom3}$ for $s=0$ and with $G$'s hierarchy for $s=1$. For $s\geq 2$, there emerge new kind  of hierarchies. For example, when $N=2$ and $s=2$, the number of moments is $15$ adopting the following moment  in addition to the $14$ moments appearing in \eqref{ET14poly}
\begin{align} 
 H^*{}^2 = 2\inta f \,  \xi^2 \, \left(\frac{m \xi^2}{2} + 2\II\right)  \,  \da . \label{mom15}
\end{align}
The theory with $15$ moments with the new moment \eqref{mom15} was the subject of the paper  \cite{Fluids-2020}. 

A claim of \eqref{relRETold} (\eqref{Hsold} for classical case) is  that the integrand of the moments is the sum of the rest energy $mc^2$ (kinetic energy $m\xi^2/2$) and $n \II$ ($s\II$) not the internal energy $\II$.  
This fact seems unphysical because we expect that we have the full energy at molecular level, i.e., $mc^2 +\II$ in relativistic context and $m \xi^2/2 + \II$ in the classical framework.

 To avoid this unphysical situation, Pennisi first noticed that $mc^2(mc^2 +2 \II)$ appearing in \eqref{PS3}$_3$ are the first two term of $(mc^2 + \II)^2$ and the same is for what concerns in \eqref{relRETold}  $(mc^2)^{n-1}(mc^2 +n \II)$  that are the first two terms of
 $(mc^2 + \II)^n$.
 Therefore he proposed in \cite{Pennisi-2021} to modify, in the relativistic case, the definition of the moments by using the substitution:
 \begin{equation}\label{Rsub}
 (mc^2)^{n-1}\left( mc^2 + n \II \right) \qquad \text{with \quad } \left( mc^2 + \II \right)^n,
 \end{equation}
 i.e., instead of \eqref{relRETold}, the following moments are proposed:
\begin{align} \label{relRET}
\begin{split}
& A^{\alpha \alpha_1 \cdots \alpha_n  } = \left(\frac{1}{mc}\right)^{2n-1} \int_{\R^{3}}
\int_0^{+\infty} f  \,  p^\alpha p^{\alpha_1} \cdots p^{\alpha_n}  \, \left( mc^2 +  \II \right)^n\, 
\phi(\mathcal{I}) \, d \mathcal{I} \, d \boldsymbol{P}  \, , \\
& I^{\alpha_1 \cdots \alpha_n  } = \left(\frac{1}{mc}\right)^{2n-1} \int_{\R^{3}}
\int_0^{+\infty} Q  \,  p^{\alpha_1} \cdots p^{\alpha_n}  \, \left( mc^2 +  \II \right)^n\, 
\phi(\mathcal{I}) \, d \mathcal{I} \, d \boldsymbol{P}. \\
\end{split}
\end{align}
This is more physical because now the full energy appears in the moments and moreover this does not modify the relativistic theory of ET$_{14}$ studied in \cite{Annals}. Instead, if we take the full triple tensor $A^{\alpha\beta\gamma}$ in \eqref{Annals}, we have a new theory with $15$ fields that is the subject of a paper in preparation by the present authors.

\bigskip

In the following sections, we construct a classical RET theory where, in analogy with \eqref{Rsub}, we modify the classical moments substituting 
\begin{equation}\label{CSt}
\left(\frac{m\xi^2}{2}\right)^{n-1}\left( \frac{m\xi^2}{2}+ n \II \right) \qquad \text{with \quad } \left( \frac{m\xi^2}{2}+ \II \right)^n.
\end{equation}
In an incoming paper, we will prove that this modifications coincide with the classical limit of the new relativistic theory with the moments given in \eqref{relRET}. 

The plan of the paper is:   in  Sect. \ref{sct3}, we propose the new hierarchies of moment equations and 
in Sect. \ref{sct4}, as a simple case, we study the theory with $15$ moments.


\section{New moment hierarchy  for polyatomic gas} \label{sct3}

As it has been discussed in the previous section, we may consider the moments such that the full energy appears at molecular level, i.e., the sum of the microscopic kinetic energy $m\xi^2 /2$ and internal energy $\II$. For this reason and according with  the previous classical limit of polyatomic gas \eqref{sysgenold} with \eqref{Hsold} and the new assumption \eqref{CSt}, we assume the following moment equations as more suitable physical moments for polyatomic gases:
\begin{align}
 \begin{split}
 &\partial_t H^s_{i_1 \cdots i_h} + \partial_i H^s_{ii_1\cdots i_h} = J^s_{i_1\cdots i_h}  \\ \label{sysgen}
 &\text{with } \ s=0, \cdots , N, \ \text{and} \ h=0, \cdots, N-s,
 \end{split}
\end{align}
where
\begin{align}
 \begin{split}
 & H^s_{i_1 \cdots i_h} = 
  \displaystyle  m \inta f  \, \xi_{i_1}\cdots \xi_{i_h}\left(\xi^2 + \frac{2\II}{m}\right)^s \da  
, \\
 & H^s_{ii_1 \cdots i_h} = 
  \displaystyle  m \inta f \, \xi_i \, \xi_{i_1}\cdots \xi_{i_h}\left(\xi^2 + \frac{2\II}{m}\right)^s \da,\\ 
 & J^s_{i_1 \cdots i_h}   
 = \displaystyle  m \inta Q \, \xi_{i_1}\cdots \xi_{i_h}\left(\xi^2 + \frac{2\II}{m}\right)^s \da  
.
 \end{split}
\label{Hs}
\end{align}
The cases with $s=0$ and $s=1$, respectively, correspond to the previous $F$- and $G$-hierarchies defined in \eqref{mom3}, i.e.,
\begin{align*}
 H^0_{i_1 \cdots i_h} = F_{i_1 \cdots i_h}, \quad  H^1_{i_1 \cdots i_h} = G_{lli_1 \cdots i_h}.
\end{align*}
On the other hand, the hierarchies with $s\geq 2$ are newly emerged. 


We remark that, similarly with the monatomic case \eqref{LMono}, the order of the highest moment is even, i.e., $2N$, and the highest moment is scalar. This fact indicates that, in principle, the moments can be integrable.

The intrinsic (velocity independent) variables are the moments in terms of the peculiar velocity $C_i=v_i - \xi_i$ instead of $\xi_i$ as follows:
\begin{align*}
 \begin{split}
 & \hat{H}^s_{i_1 \cdots i_h} = 
  \displaystyle  m \inta f \,  C_{i_1}\cdots C_{i_h}\left(C^2 + \frac{2\II}{m}\right)^s \dA  
, \\
 & \hat{H}^s_{ii_1 \cdots i_h}
  \displaystyle  m \inta f \, C_i \, C_{i_1}\cdots C_{i_h}\left(C^2 + \frac{2\II}{m}\right)^s \dA    ,\\
 & \hat{J}^s_{i_1 \cdots i_h}   
 = \displaystyle  m \inta Q \, C_{i_1}\cdots C_{i_h}\left(C^2 + \frac{2\II}{m}\right)^s \dA  
.
 \end{split}
\end{align*}
By inserting $\xi_i=v_i + C_i$ into \eqref{Hs}, the velocity dependence of the moments is obtained as follows:
\begin{align}\label{13}
 \begin{split}
  H^{s}_{i_1 \cdots i_h} 
  &=  \sum_{p=0}^{h} \sum_{r=0}^{s} \sum_{k=p}^{s+p-r} X^{(p,r,k), \, j_1 \cdots j_{k-p}j_{k-p+1} \cdots j_k}_{i_1 \cdots i_p i_{p+1} \cdots i_h} \hat{H}^{r}_{j_1 \cdots j_{k-p} j_{k-p+1} \cdots j_{k}} , 
 \end{split}
\end{align}
with
\begin{align}\label{XXX}
 \begin{split}
      X^{(p,r,k), \, j_1 \cdots j_{k-p}j_{k-p+1} \cdots j_k}_{i_1 \cdots i_p i_{p+1} \cdots i_h} =& 
\left( \begin{matrix}
h \\
p
\end{matrix}\right) 
 \, (2)^{k-p} \, \frac{s!}{r! \, (k-p)! \, (s+p-r-k)!} \cdot\\
&\cdot \left( v^2 \right)^{s+p-r-k} \,  v_{j_1} \cdots v_{j_{k-p}} \, 
\delta_{( i_1}^{j_{k-p+1}} \cdots \delta_{i_p}^{j_{k}} v_{ i_{p+1}} \cdots v_{i_h )}  \, , 
\end{split}
\end{align}
Concerning the derivation of \eqref{13}, see \ref{appendixGal}. We remark that, since $0\leq r \leq s$ and $0\leq k \leq N-r$,  $ H^{s}_{i_1 \cdots i_h}$ is expressed by the velocity independent moments with the lower order tensor.

The flux \eqref{Hs}$_2$ is decomposed into the convective and the non-convective part. The velocity dependence of the non-convective flux is expressed as follows:
\begin{align}\label{14}
H^{s}_{i i_1 \cdots i_h} - v_i H^{s}_{i_1 \cdots i_h} =  \sum_{p=0}^{h} \sum_{r=0}^{s} \sum_{k=p}^{s+p-r} X^{(p,r,k), \, j_1 \cdots j_{k-p}j_{k-p+1} \cdots j_k}_{i_1 \cdots i_p i_{p+1} \cdots i_h} \hat{H}^{r}_{i j_1 \cdots j_{k-p} j_{k-p+1} \cdots j_{k}} \, .
\end{align}
Similarly, the velocity dependence of the production terms are also expressed as follows:
\begin{align}\label{15}
J^{s}_{i_1 \cdots i_h} =  \sum_{p=0}^{h} \sum_{r=0}^{s} \sum_{k=p}^{s+p-r} X^{(p,r,k), \, j_1 \cdots j_{k-p}j_{k-p+1} \cdots j_k}_{i_1 \cdots i_p i_{p+1} \cdots i_h} \hat{J}^{r}_{j_1 \cdots j_{k-p} j_{k-p+1} \cdots j_{k}} \, .
\end{align}
The results \eqref{13}, \eqref{14}, \eqref{15} with \eqref{XXX} are in perfect agreement with the general theorem on  Galilean invariance given by Ruggeri in  \cite{Ruggeri-1989} for a general balance law system.

\section{Equilibrium values of the moments}


Let us recall the equations of state of the non-polytropic gases:
 \begin{equation}
 p=p(\rho,T)=\frac{k_B}{m}\rho T, \qquad \varepsilon \equiv \varepsilon_E( T), \label{EOS}
 \end{equation}
where $p,\varepsilon_E,T$ denote as usual the equilibrium pressure, the equilibrium specific internal energy and the absolute temperature, while $k_B$ is the Boltzmann constant. The equilibrium distribution function of the gases is deduced as follows \cite{RuggeriSpiga,Ruggeri-2020RdM} :
 \begin{align}
f_E=f^{K}_E f^{I}_E \label{fE},
\end{align}
where $f^K_E$ is the Maxwellian distribution function and $f^{I}_E$ is the distribution function of the internal mode:
\begin{align}
&f^{K}_E = \frac{\rho}{m}\left(\frac{m}{2\pi k_B T}\right)^{3/2} \exp \left(- \frac{mC^2}{2 k_B T}\right), \qquad
f^{I}_E = \frac{1}{A(T)} \exp \left(- \frac{\mathcal{I}}{k_B T}\right),
 \label{fKI}
\end{align}
with the normalization factor (partition function) $A(T)$ defined by
\begin{align}\label{partitions}
 A(T) = \int_0^{+ \infty} \exp \left(- \frac{\mathcal{I}}{k_B T}\right) \varphi(\mathcal{I}) \mathrm{d}\mathcal{I},
\end{align}
where the average of the internal energy parameter $\mathcal{I}$ is made with respect to $\varphi(\mathcal{I})d\mathcal{I}$. From  \eqref{fKI}$_2$ and \eqref{partitions}, the equilibrium distribution function of internal mode satisfies 
 \begin{align}
 	\int_{0}^{+\infty} f^{I}_E  \, \varphi(\mathcal{I}) \, d \II=1. \label{fint}
 \end{align}

The specific internal energy is the moment of $f_E$ as follows:
\begin{align}
&\varepsilon = \varepsilon_E(T) = \varepsilon_E^{K}(T) + \varepsilon_E^I(T) = \frac{1}{\rho}\inta \left(\frac{mC^2}{2} + \II \right)f_E \dA,  \label{CaEqSt}
\end{align}
where $\varepsilon_E^K$ and $\varepsilon_E^I$ are the equilibrium kinetic (translational) and internal specific energies defined by
\begin{align}  \label{int1e}
\begin{split}
	&\varepsilon_E^K(T)  = \frac{1}{\rho} \inta \frac{m C^2}{2} f_E \dA  = \frac{1}{\rho} \int_{\R^3} \frac{mC^2}{2} f^K_E \,  d{\boldsymbol C}= \frac{3}{2}\frac{k_B}{m}T, \\
    &\varepsilon_E^I(T) = \frac{1}{\rho} \inta \mathcal{I} f_E \dA = \frac{1}{m}\int_0^{+\infty} \mathcal{I} f^{I}_E \,  \varphi(\II) d\mathcal{I} =  \frac{k_B}{m}T^2\frac{\mathrm{d} \log A(T)}{\mathrm{d}T}.
\end{split}
\end{align}
The identities  \eqref{int1e} are obtained by taking into account  \eqref{fE}, \eqref{fKI}, and \eqref{partitions} and by evaluating the derivative of \eqref{partitions} with respect to $T$. Therefore, if we know the partition function $A(T)$ by a statistical-mechanical analysis, we can evaluate $\varepsilon_E^I(T)$ from \eqref{int1e}$_2$ (see for more details \cite{Ruggeri-2020RdM}). Vice versa, if the  caloric equation of state is known, from \eqref{int1e}$_2$, we can evaluate the  function $A(T)$ in integral form with respect to $T$ as follows:
\begin{align*}
	A(T)=A_0 \exp \left(\frac{m}{k_B} \int_{T_0}^T \frac{\varepsilon^I_E(T')}{T'^2}dT'\right),
\end{align*}
where $A_0$ and $T_0$ are  the inessential constants. 

By adopting \eqref{fE} with \eqref{fKI}, we can evaluate the equilibrium moments as follows (see \ref{appendixEq}):
\begin{align}
\begin{split}
 \hat{H}^{s|E}_{i_1\cdots i_h} &= m \inta f_E C_{i_1}\cdots C_{i_h} \left(C^2 + \frac{2\II}{m}\right)^s \dA \\
&= \sum_{q=0}^s  \left(\begin{matrix} s \\ q \end{matrix}\right) 
\frac{2^q \rho }{h+1} (2s-2q+h+1)!! \left(\frac{k_B T}{m}\right)^{s+\frac{h}{2}} \bar{A}_q \delta_{(i_1i_2}\cdots \delta_{i_{h-1}i_h)} ,
\label{Heq}
\end{split}
\end{align}
where  $\bar{A}_r$  is the equilibrium distribution of the internal mode with respect to $\II$ and is expressed as follows:
\begin{align}
 \bar{A}_r = \int_0^{+\infty} f^I_E \left(\frac{\II}{k_B T}\right)^r \varphi(\II) d\II. \label{barI}
\end{align}
By taking the derivative of $\bar{A}_r$ with respect to $T$, we obtain the following recurrence formula:
\begin{align*}
 \bar{A}_{r+1} = T \frac{d \bar{A}_r}{dT} + (r+\bei)\bar{A}_r, \quad \bar{A}_0 =1,
\end{align*}
where 
\begin{align}\label{Bei}
 \bei = \frac{\varepsilon^I}{\frac{k_B}{m}T}.
\end{align}
Examples of the equilibrium moments are given in the following:
\begin{align}\label{momE}
\begin{split}
 &\hat{H}^{2\EE} = \frac{p}{\rho^2} \left(15+12\bei + 4\cvi + 4\bei {}^2\right),\\
  &\hat{H}^{2\EE}_{ij} = \frac{p^3}{\rho^2} \delta_{ij} \left(35+20\bei + 4\cvi + 4 \bei{}^2 \right),\\
  &\hat{H}^{3\EE}_{} =  \frac{p^3}{\rho^2} \left(105 + 90\bei +52\cvi + 24 \bei \cvi + 8T \cvi {}' + 36 \bei {}^2 + 8 \bei {}^3\right),\\
 &\hat{H}^{3\EE}_{ij} = \frac{p^4}{\rho^3} \delta_{ij} \left(315 + 210\bei+60\bei{}^2 + 76\cvi +24\bei \cvi + 8T \cvi{}' + 8 \bei{}^3\right),\\
 &\hat{H}^{4\EE} = \frac{p^4}{\rho^3} \Big(945 + 840 \bei + 648\cvi + 48\cvi{}^2 + 416 \bei \cvi + 96 \cvi \bei{}^2 + 192 T \cvi{}' \\
&\qquad \qquad \quad  + 64 T \cvi{}' \bei + 16T^2 \cvi{}'' + 360\bei{}^2 + 96\bei{}^3 + 16\bei{}^4\Big). 
\end{split}
\end{align}
where 
\begin{align}\label{cvi}
\hat{c}_v^I =  \frac{mc_v^I}{k_B}, \quad \text{with} \quad   c_v^I = \frac{d\varepsilon^I_E(T)}{dT}
\end{align}
being the specific heat of the internal mode, and $\cvi{}' = d\cvi/dT$ and $\cvi{}'' = d^2 \cvi/dT^2$. We emphasize that $\hat{c}_v^I>0$.

\section{New ET$_{15}$ for polyatomic gases} \label{sct4}

Let us study the system \eqref{sysgen} for a given $N$. When $N=1$, the system is the Euler system. When $N=2$,  \eqref{sysgen} reduces to  the $15$ moments as follows:
\begin{align}
\begin{split}
 &\frac{\partial F}{\partial t}+\frac{\partial F_k}{\partial x_k}=0,\\
 &\frac{\partial F_i}{\partial t}+\frac{\partial F_{ik}}{\partial x_k}=0, \\
 &\frac{\partial F_{ij}}{\partial t}+\frac{\partial F_{ijk}}{\partial x_k}=P_{ij}, \ \ \ \ \ \ \ \ \ \ \frac{\partial G_{ll}}{\partial t}+\frac{\partial G_{llk}}{\partial x_k}=0,\\
 &\qquad \qquad \qquad \qquad \qquad \quad   \frac{\partial G_{lli}}{\partial t}+\frac{\partial G_{llik}}{\partial x_k}=Q_{lli},\\
  &\hspace{7.5cm}\frac{\partial H^{2}}{\partial t}+\frac{\partial H^2_{k}}{\partial x_k}=J^{2}.
\end{split}\label{balanceFG1}
\end{align}
For later convenience, we summarize the macroscopic quantities in \eqref{balanceFG1} defined as the moments of  $f$ as follows (see \eqref{sysgen}, \eqref{Hs}):
\begin{align}
 &\left( \begin{array}{l}
  F\\ F_i\\ F_{ij}  \\ F_{ijk}
 \end{array}\right)
 =
 m \inta f \, 
  \left( \begin{array}{c}
  1 \\ \xi_i\\ \xi_i \xi_j \\ \xi_i \xi_j \xi_k
 \end{array}\right)
  \da ,  \nonumber\\
  &\left( \begin{array}{l}
  G_{ll}\\  G_{lli} \\ G_{llik}
 \end{array}\right)
 =
m \inta  f \, \left(\xi^2 + 2\frac{\mathcal{I}}{m}\right)
  \left( \begin{array}{c}
  1 \\ \xi_i \\ \xi_i \xi_k
\end{array}\right)
   \da, \label{moments} \\ 
  &\left( \begin{array}{l}
  H^2\\  H^2_{i}
\end{array}\right)
  = m
 \inta f \,  \left(\xi^2 + 2\frac{\mathcal{I}}{m}\right)^2 
  \left( \begin{array}{c}
  1 \\ \xi_i
\end{array}\right)
   \da , \nonumber
\end{align}
and the production terms
\begin{align}\label{productioni}
 \mathbf{f}  \equiv &\left( \begin{array}{l}
  P_{ij} \\ Q_{lli} \\ J^{2}
\end{array}\right)
 = m
 \inta Q(f)  \,
  \left( \begin{array}{c}
  m\xi_i \xi_j \\  m\left(\xi^2 +2\frac{\mathcal{I}}{m}\right)\xi_i \\ m\left(\xi^2 + 2\frac{\mathcal{I}}{m}\right)^2 
\end{array}\right)
 \da .
\end{align}

The velocity dependence of the moments can be deduced from  \eqref{13} in the present case $N=2$, and it is obtained as follows:
\begin{align*} 
 \begin{split}
  & F=\rho , \\
  &F_i=\rho v_i, \\
  & F_{ ij} = \hat{F}_{ij} + \rho v_{ i}v_{j},  \\
  & G_{ll}= \hat{G}_{ll}+ \rho v^2 , \\
  & G_{lli}  = \hat{G}_{lli} + \hat{G}_{ll}v_i + 2\hat{F}_{li}v_{l}  + \rho v^2 v_i,\\
  & H^2 = \hat{H}^{2} + 4 \hat{G}_{lli}v_i + 2 \hat{G}_{ll}v^2 + 4\hat{F}_{ij}v_iv_j  + \rho v^4.
 \end{split}
\end{align*}
Similarly, the velocity dependences of the fluxes \eqref{14}  and productions \eqref{15} are obtained as follows:
\begin{align*}
  &F_{ijk} = \hat{F}_{ijk} + \hat{F}_{ij}v_{k} + \hat{F}_{jk}v_{i}+\hat{F}_{ki}v_{j} + \rho v_i v_j v_k ,\nonumber\\
  &G_{llik} = \hat{G}_{llik} + \hat{G}_{lli}v_{k} + \hat{G}_{llk}v_{i}+ 2\hat{F}_{lik}v_l+2\hat{F}_{kl}v_{l}v_{i}+2\hat{F}_{il}v_{l}v_{k} + \hat{F}_{ik}v^2 + \hat{G}_{ll}v_iv_k + \rho v^2 v_i v_k , \nonumber\\
  &H^2_{k} = \hat{H}^2_{k} + \hat{H}^{2}v_k + 4\hat{G}_{llik}v_i + 2 \hat{G}_{llk}v^2 + 4\hat{G}_{lli}v_iv_k + 4\hat{F}_{ijk}v_i v_j + 2\hat{G}_{ll}v^2 v_k + 4\hat{F}_{ik}v^2 v_i + 4\hat{F}_{ij}v_iv_jv_k + \rho v^4 v_k,\nonumber\\
   &P_{ij}=\hat{P}_{ij},   \\
 &Q_{lli}=2v_l \hat{P}_{il}+\hat{Q}_{lli}, \nonumber\\
  &J^{2}=4v_iv_j\hat{P}_{ij} + 4v_i \hat{Q}_{lli}+\hat{J}^{2}. \nonumber
\end{align*}

Besides $\hat{H}^2$, the velocity independent part of the moments are related to the following conventional fields:
\begin{align}
 &\text{mass density}: && \rho = \inta m f \, \dA  = \inta m f_E  \, \dA, \nonumber \\
  &\text{velocity}: && v_i = \frac{1}{\rho}\inta m f \, \xi_i  \, \da = \frac{1}{\rho}\inta m f_E \,  \xi_i  \, \da  , \nonumber\\
  &\text{specific internal energy density}: && \varepsilon =  \varepsilon^K + \varepsilon^I = \frac{1}{\rho}\inta f\,  \left(\frac{mC^2}{2}+\II\right)  \, \dA ,    \nonumber \\
 &\text{specific translational energy density}: && \varepsilon^K = \frac{1}{\rho}\inta f \, \frac{mC^2}{2}   \, \dA, \nonumber\\
 &\text{specific internal energy density}: && \varepsilon^I = \frac{1}{\rho}\inta f \, \mathcal{I}  \, \dA , \label{conventional} \\ 
  &\text{total nonequilibrium pressure}: && \PP = \frac{2}{3} \rho \varepsilon^K = \frac{1}{3}\inta  m f \, C^2  \, \dA, \nonumber\\ 
   &\text{dynamic pressure}: && \Pi = \PP - p= \frac{1}{3}\inta m \, (f-f_E) \, C^2  \, \dA , \nonumber \\
   &\text{shear stress}: && \sigma_{\langle ij\rangle} = - \inta m f \, C_{\langle i} C_{j\rangle}\,     \dA , \nonumber \\
    &\text{heat flux}: && q_i = \frac{1}{2} \inta m f \, \left(C^2 + 2\frac{\mathcal{I}}{m} \right)C_i \,  \dA , \nonumber
\end{align}
 and these are related to the intrinsic moments as follows:
\begin{align*}
 \hat{G}_{ll} = 2\rho \varepsilon  = 2 \rho (\varepsilon^K + \varepsilon^I), \quad \hat{F}_{ll} = 3\PP=3 (p+\Pi), \quad \hat{F}_{\langle ij\rangle} = -\sigma_{\langle ij\rangle}, \quad \hat{G}_{lli}= 2q_i, 
\end{align*}
where the temperature of the system $T$ is introduced through the caloric equation of state
\begin{align}
	\varepsilon =\varepsilon_E (T). \label{calEOS}
\end{align}
\noindent{\bf Remark}: As the mass density, momentum and energy density are equilibrium variables, we have in \eqref{conventional}$_{1,2,3}$ that in the moments  we can put $f$ or $f_E$ indifferently and therefore concerning energy we have:
\begin{align}\label{Poly:TotalEnergy}
\varepsilon = \varepsilon^{K}  + \varepsilon^{I} = \varepsilon^{K}_E + \varepsilon^{I}_E,
\end{align}
where $\varepsilon^{K}$ and $\varepsilon^{I}$ are defined in \eqref{conventional}$_{4,5}$ and therefore are nonequilibrium variables that are not equal to $\varepsilon^{K}_E(T)$ and $\varepsilon^{I}_E(T)$, respectively. The same   concerns the  nonequilibrium pressure ${\cal{P}}$ that is not equal to the equilibrium pressure $p$ at temperature $T$ due to the non-zero dynamic pressure $\Pi$ (see \eqref{conventional}$_{6,7}$). 


Let us decompose the intrinsic part of $H^2$ into the equilibrium part shown in \eqref{momE} and the nonequilibrium part $\Delta$ as follows:
\begin{align*}
	 \hat{H}^{2}=  \inta m \left( C^2 + 2\frac{\mathcal{I}}{m} \right)^2 f \, \dA = \frac{p^2}{\rho}(15+12\bei + 4\bei{}^2 + 4\cvi) + \Delta,
\end{align*}
where  $\Delta$ is defined by
\begin{align}
 \Delta =  \inta m \left(C^2 + 2\frac{\mathcal{I}}{m} \right)^2 (f-f_E) \, \dA. \label{defDelta}
\end{align}

The constitutive quantities are now the following moments:
\begin{align*}
   &\hat{F}_{ijk}  = \inta m C_i C_j C_k \, f \, \dA , \\
 &\hat{G}_{llik}  = \inta m\left(C^2+2\frac{\mathcal{I}}{m}\right)C_iC_k \,  f \, \dA , \\
 &\hat{H}^2_{k} = \inta m\left(C^2 + 2\frac{\mathcal{I}}{m}\right)^2 C_k \, f \, \dA,
\end{align*}
that are needed to be determined for the closure of the differential system together with the production terms $P_{ij}, Q_{lli}$ and $ J^{2}$.

\subsection{Nonequilibrium distribution function derived from MEP}
To close the system \eqref{balanceFG1}, we need the nonequilibrium distribution function $f$, which is derived from the MEP. According to the principle, the most suitable distribution function $f$ of the truncated system \eqref{balanceFG1} is the one that maximizes the entropy density
\begin{align*}
 h =&  -k_B\int_{\R^3} \int_0^{+\infty }f \log f\, \varphi(\mathcal{I})\, d\mathcal{I} d\cc , 
\end{align*}
 under the constraints that the density  moments $F, F_i, F_{ ij},  G_{ll}, G_{lli}, H^2$ 
  are prescribed as  in \eqref{moments} \cite{Dreyer,ET,RET}. Therefore, the best-approximated distribution function $f_{15}$ is obtained as the solution of an unconstrained maximum  of 
\begin{align*} 
 \mathcal{L}\left(f\right)  = &- k_B  \inta f \log f\, \da \nonumber\\
 &+ \lambda \left(F - \inta m  f \da\right)  + \lambda_i \left(F_i - \inta m \, f \, \xi_i \da\right)  \\
  & + \lambda_{ij} \left(F_{ij} - \inta m \, f \, \xi_i \xi_j \da\right)
 + {\mu} \left( G_{ll} - \inta  m \, f\, \left(\xi^2 + 2 \frac{\mathcal{I}}{m}\right)   \da \right)\nonumber\\
 & + \mu_i \left(G_{lli} - \inta  m \, f \, \left(\xi^2 + 2\frac{\mathcal{I}}{m} \right) \xi_i   \da\right)
 + {\zeta} \left( H^2 - \inta m\, f \, \left(\xi^2 + 2\frac{\mathcal{I}}{m}\right)^2    \da \right), \nonumber
\end{align*}
where $\lambda$, $\lambda_i$, $\lambda_{ij}$, $\mu$, $\mu_i$, and ${\zeta}$ are the corresponding Lagrange multipliers of the constraints. We obtain that the approximated distribution function $f_{15}$, which satisfies $\delta \mathcal{L}/\delta f=0$, is 
\begin{align}\label{fgen}
 \begin{split}
 &f_{15}=\exp\left(-1-\frac{m}{k_B}{\chi}\right), \qquad \text{with}\\
 &{\chi} = {\lambda}+\xi_i {\lambda}_i + \xi_i\xi_j{\lambda}_{ij} + \left(\xi^2 + \frac{2\mathcal{I}}{m}\right){\mu}+ \left(\xi^2+\frac{2\mathcal{I}}{m}\right)\xi_i{\mu}_i + \left(\xi^2+\frac{2\mathcal{I}}{m}\right)^2{\zeta}.
 \end{split}
\end{align}
 As $f$ is a scalar independent of frame, we have $\chi = \hat{\chi}$ where the hat indicate the same quantity evaluated in the rest frame $v_i=0$.  In this way, we have the velocity dependence of the  Lagrange multipliers (according with the general theorem given in \cite{Ruggeri-1989}). We remark that the Lagrange multipliers as fields symmetrize the system \eqref{balanceFG1}, and are called as \emph{main field} \cite{Boillat-1997,RS,newbook}. For later convenience, we denote the main field as $\mathbf{u}^\prime=\{\lambda, \lambda_i, \lambda_{ij}, \mu, \mu_i , \zeta\}$ and its velocity independent part as $\hat{\mathbf{u}}^\prime = \{ \hat{\lambda}, \hat{\lambda}_i, \hat{\lambda}_{ij}, \hat{\mu}, \hat{\mu}_i, \hat{\zeta}\}$.

Recalling the usual thermodynamical definition of the equilibrium as 
the state in  which the entropy production
 vanishes and hence attains its minimum
value, it is possible to prove the theorem \cite{BoillatRuggeri-1998,BoillatRuggeriARMA} that the components of the Lagrange multipliers of the balance laws of nonequilibrium variables vanish, and only 
the five Lagrange multipliers corresponding to the conservation laws (Euler System) remain.  On the other hand, in \cite{Pavic-2013}, it was proved that the distribution function maximizes the entropy density with the constraints of $5$ moments $F, F_i$ and $G_{ll}$ of the equilibrium subsystem is given by \eqref{fE}. Therefore, in equilibrium, $f_{15}$ coincides with the equilibrium distribution function \eqref{fE} with Lagrange multipliers given by
\begin{align}
\begin{split}
& {\lambda}_E = \frac{1}{T}\left(- g+ \frac{v^2}{2}\right), \quad {\lambda}_{i_E}= - \frac{v_i}{T}, \quad \mu_E = \frac{1}{2T}, \\
&  {{\lambda}_{\langle ij\rangle_E}} =0 , \quad {\lambda}_{ll_E} = 0, \quad {{\mu}_{i_E}}=0, \quad  \zeta_E=0,
\end{split} \label{mainE}
\end{align}
where $g =\varepsilon_E(T) +p/\rho -T s$ is the equilibrium chemical potential. We remark that ${\lambda}_E , {\lambda}_{i_E},  \mu_E$ in \eqref{mainE}   are the \emph{main field} that symmetrize the Euler system as was proved first by Godunov (see \cite{Godunov,book}).

When we adopt \eqref{fgen} to derive the closed constitutive equations, we need to take care of the Junk's problem \cite{Junk-1998} that the domain of definition of the flux in the last moment equation is not convex, the flux has a singularity, and the equilibrium state lies on the border of the domain of definition of the flux. To avoid these difficulties in the molecular ET approach,  we consider, as usual, the processes near equilibrium.  For this reason, we expand \eqref{fgen} around an equilibrium state as follows:
\begin{align}\label{fgenE}
 \begin{split}
 &f_{15} \simeq f_E\left(1-\frac{m}{k_B}\tilde{\chi}\right), \\
 &\tilde{\chi} = \tilde{\lambda}+C_i \tilde{\lambda}_i + C_iC_j\tilde{\lambda}_{ij} + \left(C^2 +\frac{2\mathcal{I}}{m}\right)\tilde{\mu} + \left(C^2+\frac{2\mathcal{I}}{m}\right)C_i\tilde{\mu}_i + \left(C^2+\frac{2\mathcal{I}}{m}\right)^2 \tilde{\zeta},
 \end{split}
\end{align}
where a tilde on a quantity indicates its velocity independent nonequilibrium part. In the following, for simplicity, we use the notation $f$ instead of  $f_{15}$. 


 Inserting \eqref{fgenE} into \eqref{conventional} and \eqref{defDelta}, we obtain the following algebraic relation for Lagrange multipliers:
\begin{align}
\begin{split}
 &\left(\begin{array}{c}
  0 \\ -3 \frac{k_B}{m}\Pi \\ 0 \\ - \frac{k_B}{m} \Delta \\
       \end{array}\right)
 =
  \left(\begin{array}{cccc}
   \hat{H}^{0\EE} & \hat{H}_{ll}^{0\EE} & \hat{H}^{1\EE} & \hat{H}^{2\EE} \\
   \hat{H}_{ll}^{0\EE} & \hat{H}_{llmm}^{0\EE} & \hat{H}_{ll}^{1\EE} & \hat{H}_{ll}^{2\EE} \\
   \hat{H}^{1\EE} & \hat{H}_{ll}^{1\EE} & \hat{H}^{2\EE} & \hat{H}^{3\EE} \\
   \hat{H}^{2\EE} & \hat{H}_{ll}^{2\EE} & \hat{H}^{3\EE} & \hat{H}^{4\EE} \\
  \end{array}\right)
 \left(\begin{array}{c}
  \tilde{\lambda} \\ \frac{1}{3}\tilde{\lambda}_{ll} \\ \tilde{\mu} \\ \tilde{\zeta} \\
       \end{array}\right), \\
 &\left(\begin{array}{c}
  0 \\ -6 \frac{k_B}{m}q_i  \\
       \end{array}\right)
 =
  \left(\begin{array}{cc}
   \hat{H}^{0\EE} & \hat{H}_{ll}^{0\EE}  \\
   \hat{H}_{ll}^{0\EE} & \hat{H}_{llmm}^{0\EE} \\
  \end{array}\right)
 \left(\begin{array}{c}
  \tilde{\lambda}_i \\ \tilde{\mu}_i \\
       \end{array}\right), \\
 &\text{and} \quad
 \frac{k_B}{m}\sigma_{\langle ij\rangle} = \hat{H}^{0\EE}_{\langle ij\rangle \langle rs\rangle}\tilde{\lambda}_{\langle rs\rangle}. 
\end{split}
\label{Lags}
\end{align}
Taking into account \eqref{momE}, as the solutions of \eqref{Lags}, the nonequilibrium parts of the Lagrange multipliers are obtained as follows:
\begin{align}
\begin{split}
 &\tilde{\lambda} = \frac{1}{8\cvi \rho T}\left\{\tilde{\Pi}\left(4\cvi {}^2 - 4T \cvi {}' \bei - \cvi(15+20\bei+4\bei {}^2)\right) + 12\Pi \left(\cvi - \bei\right)\right\},\\
 &\tilde{\lambda}_{ll} = - \frac{3}{4\cvi p T}\left\{T \cvi {}' \tilde{\Pi} + (3+2\cvi) \Pi \right\} ,\\
 &\tilde{\mu} =   \frac{1}{4\cvi p T}\left\{\left(\cvi(2\bei + 5)+T \cvi {}' \right)\tilde{\Pi} + 3 \Pi \right\} ,\\
 &\tilde{\zeta} = - \frac{\rho}{8p^2 T}  \tilde{\Pi}, \\
 &\tilde{\lambda}_i = \frac{2\bei +5}{(2\cvi+5)p T}q_i,\\ 
 &\tilde{\mu}_i = - \frac{\rho}{(2\cvi + 5)p^2 T }q_i, \\
 &\tilde{\lambda}_{\langle ij\rangle} = \frac{1}{2pT} \sigma_{\langle ij\rangle},
 \end{split}
\label{Lagrange}
\end{align}
where $\tilde{\Pi}$ is the nonequilibrium variable introduced instead of $\Delta$ defined by
\begin{align}
 \tilde{\Pi} = \frac{1 }{R}\left(\frac{\rho}{p}\Delta + 6\frac{T\cvi {}'}{\cvi}\Pi\right),\label{tilpi}
\end{align}
with
\begin{align}
 R =   (2\cvi +3)(2\cvi +5) +4 T \cvi {}'  -2T^2 \frac{\cvi {}'{}^2}{\cvi} + 2 T^2 \cvi {}''. \label{R}
\end{align}

 \subsection{Constitutive equations}

By using the distribution function \eqref{fgenE} with \eqref{Lagrange}, we obtain the constitutive equations for the fluxes up to the first order with respect to the nonequilibrium variables as follows:
\begin{align*}
 \begin{split}
   &\hat{F}_{ijk}  
 = \frac{2}{2\hat{c}_v^I +5} (q_k \delta_{ij} + q_j \delta_{ik} + q_i \delta_{jk}),\\
 &\hat{G}_{llij}  
 = (2\bei+5)\frac{p^2}{\rho}\delta_{ij} + (2\bei + 7)\frac{p}{\rho}\Pi \delta_{ij} + (2\hat{c}_v^I +5)\frac{p}{\rho} \tilde{\Pi}  \delta_{ij} - (2\bei + 7)\frac{p}{\rho} \sigma_{\langle ij\rangle},\\
 &\hat{H}^2_{i} 
 = \frac{4}{2\hat{c}_v^I + 5} \frac{p}{\rho} \left(35+2T \cvi{}' + 10 \bei + 2\cvi (2\bei+7)\right)q_i.  
 \end{split}
\end{align*}

\subsection{Entropy density and flux}

The entropy density $h$ satisfies the entropy balance equation:
\begin{align}
 \frac{\partial h}{\partial t} + \frac{\partial}{\partial x_i} (hv_i + \varphi_i) = \Sigma,
\label{entropyineq}
\end{align}
where $\varphi_i$ and $\Sigma$ are, respectively, the non-convective entropy flux defined below and the entropy production studied in \eqref{Sigma}.

By adopting \eqref{fgenE} with \eqref{Lagrange}, we obtain the entropy density within the second order with respect to the nonequilibrium variables  as follows:
\begin{align}\label{entropia}
 \begin{split}
 h 
 =&\rho s  - \frac{3(2\hat{c}_v^I+3)}{8\hat{c}_v^I p T} \Pi^2 
  - \frac{R}{16pT}\tilde{\Pi}^2
 -\frac{1}{4pT}\sigma_{\langle ij\rangle}\sigma_{\langle ij\rangle} - \frac{\rho}{(2\hat{c}_v^I+5)p^2 T} q_i q_i.
 \end{split}
\end{align}
This means that the entropy density is convex and reaches the maximum at equilibrium. When $R>0$, the entropy density is convex and then the system \eqref{balancekc} provides the symmetric form in the main field components.  Concerning the condition $R>0$, we will discuss  in Sect. \ref{secR}.

Similarly, the entropy flux is obtained as follows:
\begin{align*}
 \varphi_i  =& -k\int_{\R^3} \int_0^{+\infty} C_i f \log f\, \dA\\ 
 = & \frac{1}{T}q_i + \frac{2}{pT (2\hat{c}_v^I +5)}q_j \sigma_{\langle ij\rangle}  - \frac{2}{pT (2\hat{c}_v^I +5)}q_i \Pi - \frac{1}{pT}q_i \tilde{\Pi},
\end{align*}



\subsection{Production terms with BGK collisional model}

In the present paper, we evaluate the production terms, for simplicity, with the usual BGK model:
\begin{align}\label{bgkk}
 Q(f) = -   \frac{1}{\tau}(f-f_E).
\end{align}
%
%
 Taking into account  \eqref{productioni} and  \eqref{bgkk} we have:
\begin{align}
 \begin{split}
 &\hat{P}_{ll} = - \frac{3}{\tau}\Pi, \quad \hat{P}_{\langle ij\rangle} =    \frac{1}{\tau} \sigma_{\langle ij\rangle}, \quad 
 \hat{Q}_{lli} = -   \frac{2}{\tau} \, q_i,  
 \quad \hat{J}^2 =  - \frac{1}{\tau}\Delta .
 \end{split}
\label{prods}
\end{align}

According to the symmetrization theorem \cite{RS,book,newbook}, the entropy production $\Sigma$  defined in \eqref{entropyineq} is obtained as the scalar product between the main field $\mathbf{u}^\prime$ and the production vector $\mathbf{f}$ given by \eqref{productioni}. Moreover, from the consideration on the velocity dependence of the fields studied in \cite{Ruggeri-1989}, the production is same with the scalar product between the velocity independent part of the main field $\hat{\mathbf{u}}^\prime$ and the velocity independent part of the production vector $\hat{\mathbf{f}}$. In conclusion, by taking into account \eqref{Lagrange} and \eqref{prods},  we have
\begin{align}
\begin{split}
 \Sigma = & \mathbf{u} ^\prime \cdot \mathbf{f}  = \hat{\Sigma} =  \hat{\mathbf{u}} ^\prime \cdot \hat{\mathbf{f} } = \frac{\hat{\lambda}_{ll}}{3}\frac{\Pi}{\tau  }  - \hat{\lambda}_{\langle ij\rangle}\frac{\sigma_{\langle ij\rangle}}{\tau  } + 2 \hat{\mu}_i \frac{q_i}{\tau  } + \hat{\zeta}\frac{\Delta}{\tau  }\\
  =& \frac{3(2\hat{c}_v^I +3)}{4\hat{c}_v^I pT}\frac{1}{\tau  }  {\Pi}^2 + \frac{R}{8pT \tau}\tilde{\Pi}^2
  + \frac{1}{2pT} \frac{1}{\tau} \, \sigma_{\langle ij\rangle}\sigma_{\langle ij\rangle} + \frac{2\rho}{p^2T(2\hat{c}_v^I+5)} \frac{1}{\tau}\, q_i q_i.  \label{Sigma}
\end{split}
\end{align}
Under the condition that $R>0$, we notice that $\Sigma \geq0$.

In order to evaluate more precisely the production terms, a generalized BGK with two relaxation times is used in the literature \cite{Struchtrup-1999,Struchtrup-2014} (see also \cite{Ruggeri-2020RdM,ET7,ET15}). 

\subsection{Estimation of $R$: convexity of entropy density and positivity of entropy production}\label{secR}

As we have seen in \eqref{entropia} and \eqref{Sigma}, if $R$ defined in \eqref{R} is positive, the entropy density is convex and the entropy production is positive. 
To estimate $R$, we need the dependence of the specific heat on the temperature.
\begin{figure}[htbp]
  \begin{center}
   \includegraphics[width=56mm]{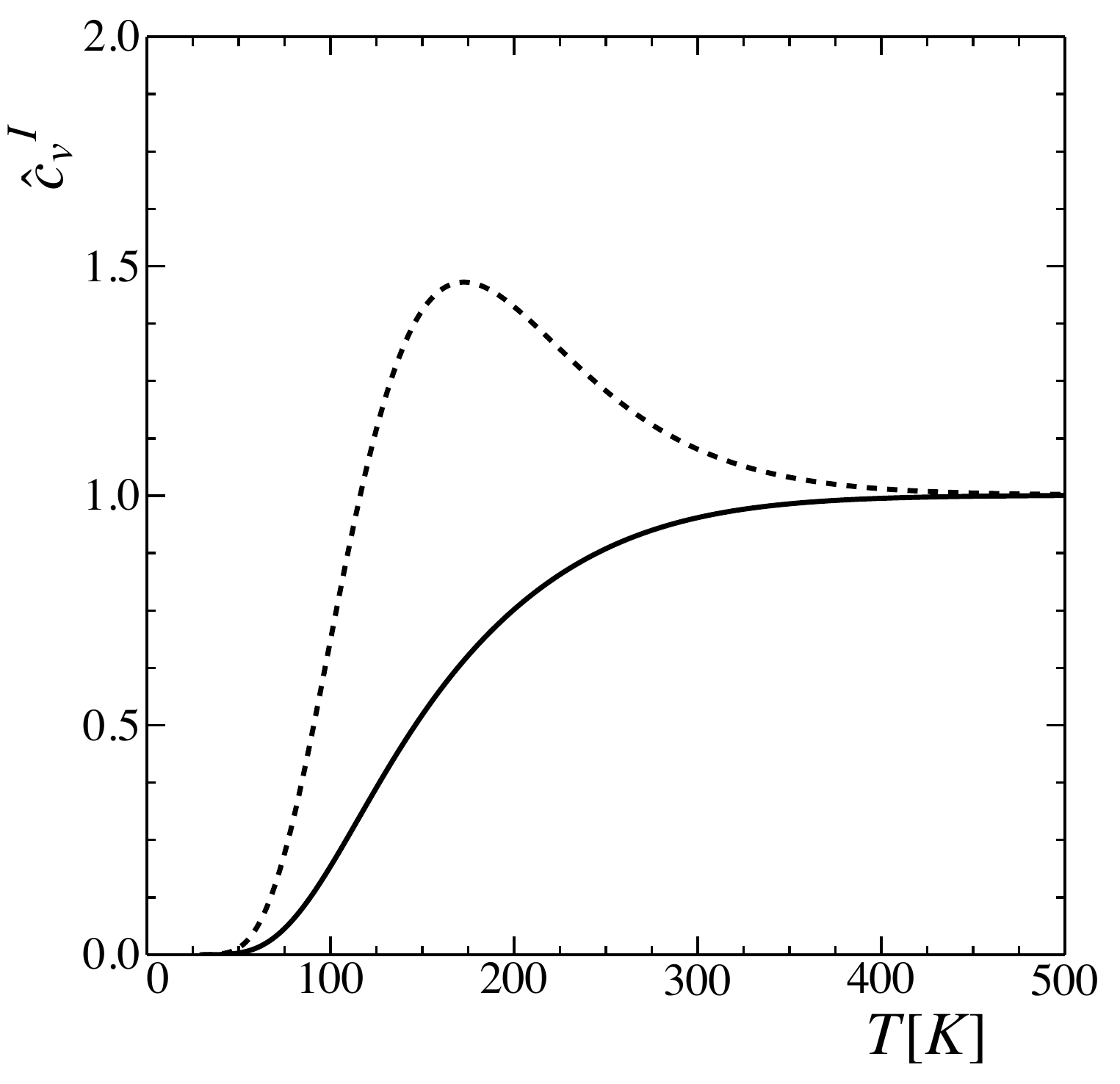}
   \includegraphics[width=56mm]{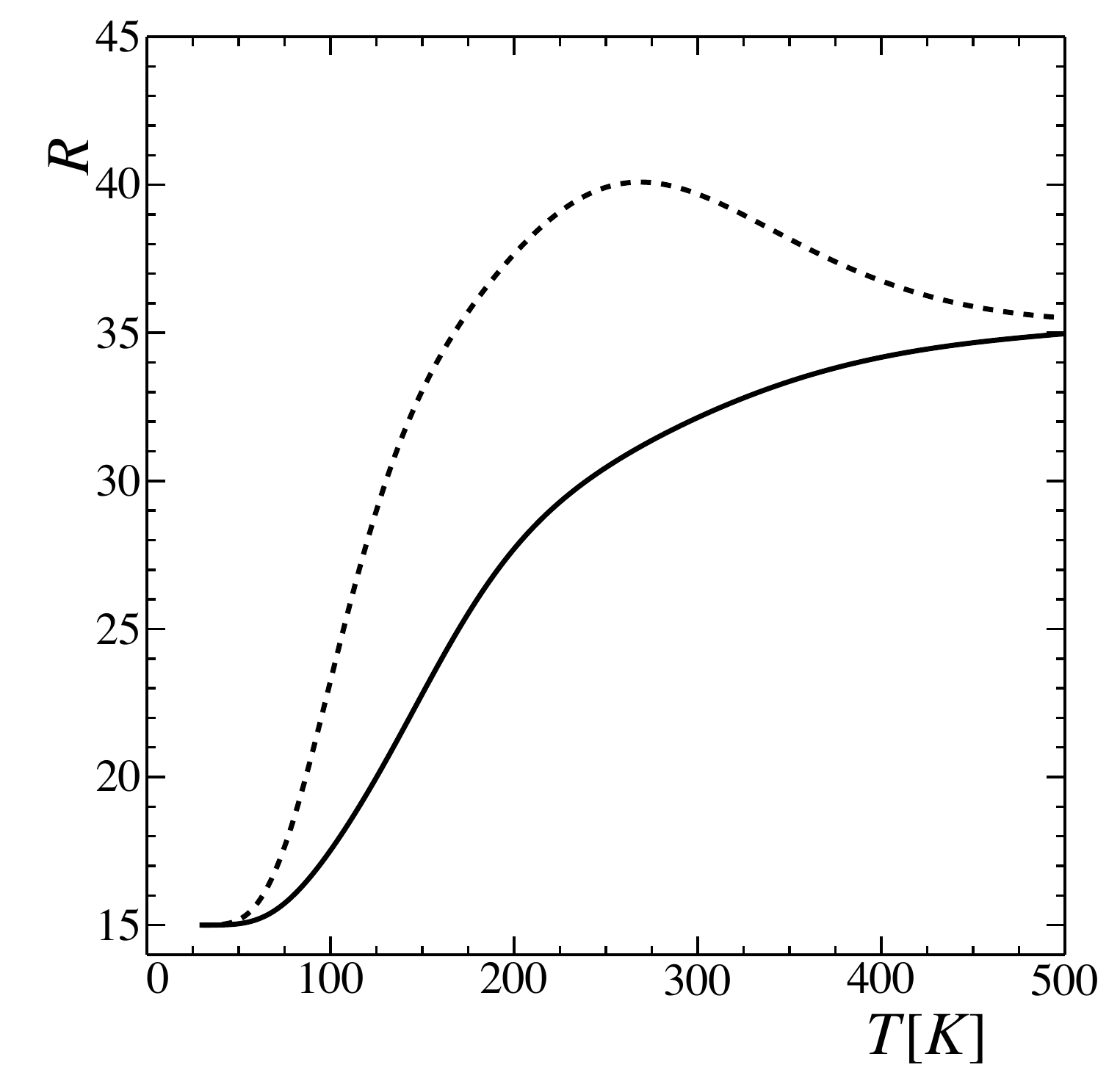}
  \end{center}
  \caption{Dependence of the dimensionless specific heat of internal mode $\hat{c}_v^I$ (left) and $R$ (right) for normal-Hydrogen (solid line) and para-Hydrogen (dashed line) on the temperature $T$ in the temperature range from $30$K to $500$K.}
  \label{fig-cv}
\end{figure}
As an example, let us consider the case of normal hydrogen and para hydrogen gases in a low temperature range where we can safely neglect the contribution of the vibrational mode. 
 The reason of the choice of the gases are the following. First, since the value of the specific heat of the hydrogen gases is small, the value of $R$ is small. Second, the temperature dependence of $\hat{c}_v^I$ is important in the estimation of $R$. As we will see in Fig. \ref{fig-cv}, the temperature dependence of normal hydrogen is monotonic as usual gases and the one of para hydrogen has a peak at a temperature. 

The specific heat of the internal mode is estimated on the basis of statistical mechanics \cite{LandauSM,LandauQM} as follows:
\begin{align*}
 &\hat{c}_{v}^I = \beta^2 \frac{\partial^2 \log Z_{\mathrm{rot}}}{\partial \beta^2},
  \ \ \ \left(\beta \equiv  \frac{1}{k_B T}\right)    
\end{align*}
where $Z_{\mathrm{rot}}$ is the partition function due to the rotational modes. The partition function is given by 
\begin{align*}
 \begin{split}
 &Z_{\mathrm{rot}}=Z_g^{g_g}Z_u^{g_u}, \\
 &Z_{g} = \sum_{l=even}(2l+1)\exp \left[-\beta B l (l+1)\right],\\
 &Z_{u} = \sum_{l=odd}(2l+1)\exp \left[-\beta B l (l+1)\right],
  \end{split} 
\end{align*}
where $l$ is the quantum number of the orbital angular momentum and $B=\hbar^2 / 2I $ $=12.09\times 10^{-22}$ [J]\cite{Radzig} with 
$I$ and $\hbar$ being the moment of inertia of a molecule and the Planck constant divided by 2$\pi$, respectively, and $g_g$ and $g_u$ are defined by 
\begin{align*}
 \begin{split}
 &
\begin{array}{lll}
 \mathrm{normal-hydrogen} : & g_u = 3/4, \ & g_g = 1/4 \\
 \mathrm{para-hydrogen} : & g_u = 0, & g_g = 1
 \end{array} 
. \ \ 
 \end{split} 
\end{align*}

The temperature dependence of $\hat{c}_v^I$ and $R$ is shown in Fig. \ref{fig-cv}. From the figure, $R$ is positive. The situation is similar for other gases.

\section{Closed field equations}

Using the constitutive equations above, we obtain the closed system of field equations for the 15 independent fields $(\rho, v_i, T, \Pi, \sigma_{\langle ij\rangle},q_i, \Delta)$ :
\small{
\begin{align}
&\frac{\partial \rho}{\partial t}+\frac{\partial}{\partial  x_i}( \rho v_i ) = 0, \nonumber\\
 &\frac{\partial \rho v_j}{\partial t} + \frac{\partial }{\partial x_i}\left\{ [p+\Pi] \delta_{ij} -\sigma_{\langle ij\rangle}+ \rho v_i v_j\right\} = 0,  \nonumber\\
  &\frac{\partial}{\partial t} \left\{ p(2\bei +3) +\rho v^2 \right\} 
 +\frac{\partial}{\partial x_i} \left\{ 2q_i+  \left[p(2\bei + 5) +2\Pi\right]v_i - 2\sigma_{\langle li\rangle}v_l  + \rho v^2v_i\right\} =0, \nonumber \\
 &\frac{\partial}{\partial t}\left\{ 3\left(p+\Pi\right) +\rho v^2 \right\} 
 + \frac{\partial}{\partial x_k}\left\{ \frac{10}{2\hat{c}_v^I+5}q_k + 5(p+\Pi)v_k -2\sigma_{\langle lk\rangle}v_l  +  \rho v^2v_k \right\} 
= -\frac{3\Pi}{\tau  }, \nonumber\\
&\frac{\partial}{\partial t}\left(-\sigma_{\langle ij\rangle} + \rho v_{\langle i}v_{j\rangle}\right)
 +\frac{\partial}{\partial x_k}\left\{\frac{2}{1+\hat{c}_v} q_{\langle i}\delta_{j\rangle k} + 2[p+\Pi]v_{\langle i}\delta_{j\rangle k} - \sigma_{\langle ij\rangle}v_k - 2\sigma_{\langle k\langle i\rangle}v_{j\rangle} + \rho v_{\langle i}v_{j\rangle}v_k\right\} =   \frac{1}{\tau  }\,  \sigma_{\langle ij\rangle}, \nonumber\\
&\frac{\partial}{\partial t}\left\{2q_i +  \left[p(2\bei+5)+2\Pi\right]v_i -2 \sigma_{\langle li\rangle}v_l + \rho v^2 v_i\right\}+ \nonumber \\
 &\quad +  \frac{\partial}{\partial x_k} \Bigg\{ \frac{p}{\rho}\delta_{ik}\left[(2\bei+5)p + (2\bei + 7)\Pi  + (2\cvi+5)\frac{1}{R}\left(\frac{\rho}{p}\Delta + 6\frac{T\cvi{}'}{\cvi}\Pi\right)\right]
 - (2\bei + 7)\frac{p}{\rho}  \sigma_{\langle ik\rangle} \label{balancekc} \\
 &\qquad \qquad \ + \frac{4}{2\hat{c}_v^I +5}q_lv_l \delta_{ik}+ 2 \frac{2\hat{c}_v^I+7}{2\hat{c}_v^I+5}(q_iv_k + q_kv_i) 
 + (p+\Pi)v^2\delta_{ik} + \left[(2\bei+7) p + 4\Pi\right]v_iv_k \nonumber \\
 &\qquad \qquad \ - \sigma_{\langle ik\rangle}v^2 - 2 \sigma_{\langle lk\rangle}v_lv_i - 2v_lv_k \sigma_{\langle il\rangle} + \rho v^2 v_i v_k\Bigg\} = - \frac{2}{\tau  } \left( \Pi v_i      -  {\sigma_{\langle il\rangle}} v_l  +      q_i\right),\\
 &\frac{\partial}{\partial t}\left\{\frac{p^2}{\rho}\left(15+12\bei +4\cvi+4\bei{}^2\right)
 +\Delta
 + 8v_i q_i + 2v^2 [p(2\bei+5) + 2 \Pi] - 4v_iv_j\sigma_{\langle ij\rangle}  + \rho v^4 \right\} + \nonumber \\
 &\quad  + \frac{\partial}{\partial x_k} \Bigg\{ \frac{4}{2\cvi+5}\frac{p}{\rho}\left[35+2T\cvi {}' +10 \bei + 2\cvi(2\bei+7)\right]q_k 
+\frac{p^2}{\rho}\left(35+20\bei +4\cvi + 4\bei{}^2\right)v_k \nonumber \\
 &\qquad \qquad \
 +\frac{p}{\rho} \left[\frac{4}{R}(2\cvi + 5) + 1\right]v_k \left(\frac{\rho}{p}\Delta + 6\frac{T\cvi{}'}{\cvi}\Pi\right)
  + \frac{p}{\rho}\left[4(2\bei+7)-6\frac{\cvi{}'}{\cvi}\right] v_k \Pi
 -4(2\bei+7)\frac{p}{\rho}v_l \sigma_{\langle lk\rangle } \nonumber\\
 &\qquad \qquad \ 
 +4\frac{2\cvi+7}{2\cvi+5}v^2 q_k + 8\frac{2\cvi+7}{2\cvi+5} v_k v_l q_l + 2\left(2\cvi+7p+4\Pi\right)v^2v_k -4 \sigma_{\langle lk\rangle}v^2v_l -4 v_kv_iv_j\sigma_{\langle ij\rangle}+\rho v^4 v_k
   \Bigg\} \nonumber \\
 &\qquad \qquad \    
 =  - \frac{1}{\tau  } \left( 4 \Pi v^2  -   4 v_{\langle i}v_{j\rangle}{\sigma_{\langle ij\rangle}} -  8  v_i {q_i}
  +  \Delta\right) . \nonumber
\end{align}
}
\normalsize
The system \eqref{balancekc} 
 is formed by $15$ equations in the $15$ unknown and  is closed   provided we assign the equilibrium equations of state \eqref{EOS} and the relaxation time. We may notice that the closed field equations of $\rho, v_i, \Pi$ and $\sigma_{\langle ij\rangle}$ are same with ET$_{14}$, and the effort of the new moment appears only in the equation of $q_i$.

The closed set of the field equations are also expressed with the material derivative denoted by a dot on a quantity such as
\begin{align*}
 \dot{f} = \frac{\partial f}{\partial t} + v_i \frac{\partial f}{\partial x_i}.
\end{align*}
Then we rewrite the closed field equations in the following form:
\small{
\begin{align}
 &\dot{\rho} + \rho \frac{\partial v_l}{\partial x_l} = 0,\nonumber \\
 &\rho \dot{v}_i + \frac{\partial p}{\partial x_i} + \frac{\partial \Pi}{\partial x_i} - \frac{\partial \sigma_{\langle ik\rangle}}{\partial x_k} =0,\nonumber \\
 &\dot{T} + \frac{2}{2\cvi+3}\frac{T}{p} \left\{(p+\Pi)\frac{\partial v_l}{\partial x_l} - \sigma_{\langle ik\rangle}\frac{\partial v_k}{\partial x_i}+\frac{\partial q_l}{\partial x_l}\right\}
 =0, \nonumber \\ 
 &\dot{\Pi} + \frac{4\hat{c}^I_v}{6\hat{c}_v^I+9} p\frac{\partial v_l}{\partial x_l} + \frac{10\hat{c}_v^I +9}{6\hat{c}_v^I + 9}\Pi \frac{\partial v_l}{\partial x_l} - \frac{4\hat{c}^I_v}{6\hat{c}_v^I+9}\sigma_{\langle lk\rangle}\frac{\partial v_l}{\partial x_k} + \frac{10}{3}q_l \frac{\partial}{\partial x_l} \left(\frac{1}{2\hat{c}_v^I +5}\right) + \frac{8\hat{c}_v^I}{3(2\hat{c}_v^I+3)(2\hat{c}_v^I+5)}\frac{\partial q_l}{\partial x_l} = - \frac{\Pi}{\tau  }, \nonumber \\
 &\dot{\sigma}_{\langle ij\rangle} + \sigma_{\langle ij\rangle}\frac{\partial v_l}{\partial x_l} + 2\sigma_{\langle l\langle i\rangle}\frac{\partial v_{j\rangle}}{\partial x_l} - 2 (p+\Pi)\frac{\partial v_{\langle j}}{\partial x_{i\rangle}} -4q_{\langle i}\frac{\partial}{\partial x_{j\rangle }} \left(\frac{1}{2\hat{c}_v^I+5}\right)  - \frac{4}{2\hat{c}_v^I+5} \frac{\partial q_{\langle i}}{\partial x_{j\rangle}} = -  \frac{1}{\tau  } \sigma_{\langle ij\rangle}, \nonumber\\
 & \dot{q}_i + \left(1+\frac{2}{2\hat{c}_v^I+5}\right)q_i \frac{\partial v_l}{\partial x_l} + \left(1+\frac{2}{2\hat{c}_v^I+5}\right) q_l \frac{\partial v_i}{\partial x_l} + \frac{2}{2\hat{c}_v^I+5}q_l \frac{\partial v_l}{\partial x_i} \nonumber \\
& \qquad + \frac{2\cvi +5}{2}\frac{p}{\rho T}\left\{(p+\Pi)\delta_{il} - \sigma_{\langle il\rangle}\right\} \frac{\partial T}{\partial x_l} - \frac{p}{\rho^2} \left(\Pi \delta_{il} - \sigma_{\langle il\rangle}\right)\frac{\partial \rho}{\partial x_l}  \label{fieldeqsM} \\
&\qquad + \frac{1}{\rho}\left\{(p-\Pi)\delta_{il}+ \sigma_{\langle il\rangle} \right\}\left(\frac{\partial \Pi}{\partial x_l}  - \frac{\partial \sigma_{\langle rl\rangle}}{\partial x_r}\right) + \frac{1}{R}\left(\frac{\rho}{p}\Delta + 6\frac{T \cvi{}'}{\cvi}\Pi\right)\frac{\partial}{\partial x_i} \left\{ \frac{p}{\rho}\left(2\hat{c}_v^I +5\right)\right\} = -  \frac{1}{\tau  }\, q_i, \nonumber \\
 & \dot{\Delta} + \left\{ \Delta  - \frac{8 T\hat{c}_v^I{}'}{2\hat{c}_v^I+3}\frac{p^2}{\rho} +  \frac{4}{R}\left(2\hat{c}_v^I +5\right)\left(\Delta + 6\frac{T \cvi{}'}{\cvi}\frac{p}{\rho}\Pi\right) + \frac{8p}{\rho(2\hat{c}_v^I+3)} \left(3+2\hat{c}_v^I -T\hat{c}_v^I{}'\right)\Pi \right\}\frac{\partial v_l}{\partial x_l} \nonumber \\
&\qquad -\frac{8p}{\rho (2\hat{c}_v^I+3)}\left(3+2\hat{c}_v^I - T\hat{c}_v^I{}'\right)\sigma_{\langle ik\rangle}\frac{\partial v_i}{\partial x_k} - \frac{8}{\rho}q_i \frac{\partial p}{\partial x_i}\nonumber \\
 & \qquad + 
 4\frac{p}{\rho T}\left\{(2\cvi+7) + \frac{2}{2\cvi+5}(2T\cvi{}' + T^2 \cvi{}'' )-\frac{4T^2 \cvi{}' {}^2}{(2\cvi+5)^2}\right\}\frac{\partial T}{\partial x_l} q_l \nonumber \\
 & \qquad + \frac{8p}{\rho}\left\{1-\frac{2T\hat{c}_v^I{}'}{(2\hat{c}_v^I+3)(2\hat{c}_v^I+5)}\right\} \frac{\partial q_l}{\partial x_l} - \frac{8}{\rho}q_i \frac{\partial \Pi}{\partial x_i} + \frac{8}{\rho} q_i \frac{\partial \sigma_{\langle il\rangle}}{\partial x_l}  =   - \frac{1}{\tau  }\Delta. \nonumber
\end{align}
}
\normalsize

\subsection{Maxwellian iteration and phenomenological coefficients}\label{maxIt}

The Navier-Stokes-Fourier theory is obtained by carrying out the Maxwellian iteration \cite{Ikenberry} on \eqref{fieldeqsM} in which only the first order terms with respect to the relaxation time are retained. The method is based on putting to zero the nonequlibrium variables on  the left side of equations \eqref{fieldeqsM}$_{4,5,6,7}$:
\begin{align}
 &\Pi = - p\tau    \frac{4\hat{c}_v^I}{6\hat{c}_v^I + 9} \frac{\partial v_k}{\partial x_k}, \quad
 \sigma_{\langle ij\rangle} = 2p \tau    \frac{\partial v_{\langle i}}{\partial x_{j \rangle}}, \quad
 q_i = - p \tau   \frac{2\hat{c}_v^I+5}{2} \frac{k_B}{m} \frac{\partial T}{\partial x_i}, \label{NFS1} 
\end{align}
and
\begin{equation} \label{deltino}
\Delta = \tau    \,   \frac{8p^2}{\rho}  \frac{T\hat{c}_v^I{}' }{{2\hat{c}_v^I+3}} \frac{\partial v_k}{\partial x_k} .
\end{equation}
Recalling the definitions of the bulk viscosity $\nu$, shear viscosity $\mu$, and heat conductivity $\kappa$ in the Navier-Stokes-Fourier theory:
\begin{align}\label{NFSs}
 \Pi =-\nu \frac{\partial v_i}{\partial x_i},
 \qquad
 \sigma_{\langle ij\rangle} = 2 \mu \frac{\partial v_{\langle i}}{\partial x_{j\rangle}}, \qquad
 q_i = - \kappa \frac{\partial T}{\partial x_i},
\end{align}
we have from \eqref{NFS1} and \eqref{NFSs}:
\begin{align}
 \begin{split}
 &\nu =  \frac{4\hat{c}_v^I}{6\hat{c}_v^I + 9} p \ \tau    ,\qquad
 \mu = p \, \tau  ,   \qquad 
 \kappa = \frac{2\hat{c}_v^I+5}{2} p \, \frac{k_B}{m}  \, \tau,  
 \end{split}
\label{maxwellite}
\end{align}
that are the same of $14$ moments \cite{newbook}. Therefore the Maxwellian Iteration of ET$_{15}$ and ET$_{14}$ give both the same parabolic  Navier-Stokes-Fourier system. 

\section{ET$_{14}$ as Principal subsystem}

Since ET$_{15}$ includes a larger set of the  equations compared to the ET$_{14}$, it is natural to expect that the ET$_{14}$ is a special case of ET$_{15}$, although the theories are based on  different entropy densities which maximize the corresponding system. In fact, ET$_{15}$ includes ET$_{14}$ as special case because it is a principal subsystem according with the definition given in \cite{BoillatRuggeriARMA}.


In fact, in the present case, the ET$_{14}$ is obtained as a principal subsystem of ET$_{15}$ under the condition $\zeta=0$, i.e., from \eqref{Lagrange}$_7$,
\begin{align*}
 \tilde{\Pi} = 0,
\end{align*}
or, in other words putting in the first $14$ equations of \eqref{fieldeqsM},
\begin{align*}
 \Delta = -6 \frac{p}{\rho}\frac{T \cvi{}'}{\cvi}\Pi ,
\end{align*}
and the  last equation of \eqref{fieldeqsM} is deleted.

\section{Polytropic Gases}

As a special case, let us consider a polytropic gas  in which $\varepsilon$ linearly depends on $T$, that is, the specific heat is a constant, and the caloric equation of state is given by 
\begin{equation*}
 \varepsilon = \frac{D}{2} \frac{k_B}{m}T, 
\end{equation*}
where the constant $D$ denotes  the internal degrees of freedom and in the monatomic gas $D=3$. In this case  we have  from \eqref{Bei} and \eqref{cvi}
\begin{align*}
\bei =         \hat{c}_v^I = \frac{D-3}{2}.
\end{align*}

Moreover,  as was obtained in \cite{Bourgat-1994,Pavic-2013},  the measure $\varphi(\II)$ is explicitly expressed with respect to  the internal degrees of freedom $D$ as follows:
\begin{align*}
 \varphi(\II) = \II^\alpha \quad \text{where} \quad \alpha = \frac{D-5}{2}.
\end{align*}
Then, the moments of the internal mode $\eqref{barI}$ is expressed simply as follows:
\begin{align}
 \bar{A}_r = \frac{\Gamma (r+\alpha+1)}{\Gamma(\alpha+1)},
\end{align}
where $\Gamma(z)$ is the gamma function.

In the present case, $R$ given in \eqref{R} becomes:
\begin{align*}
 R= D(D+2),
\end{align*}
and is positive. 
Then, from \eqref{tilpi}, the relation between $\Delta$ and $\tilde{\Pi}$ is the following:
\begin{align}\label{PiD}
 \tilde{\Pi} = \frac{\rho}{D(D+2)p}\Delta.
\end{align}

The field equations using  the material derivative \eqref{fieldeqsM} are expressed as follows (with $\tilde{\Pi}$ and $\Delta$ related by \eqref{PiD}) :
\begin{align}
\begin{split}
 &\dot{\rho} + \rho \frac{\partial v_l}{\partial x_l} = 0,\\
 &\rho \dot{v}_i + \frac{\partial p}{\partial x_i} + \frac{\partial \Pi}{\partial x_i} - \frac{\partial \sigma_{\langle ik\rangle}}{\partial x_k} =0,\\
 &\dot{T} + \frac{2T}{Dp} \left\{(p+\Pi)\frac{\partial v_l}{\partial x_l} - \sigma_{\langle ik\rangle}\frac{\partial v_k}{\partial x_i}+\frac{\partial q_l}{\partial x_l}\right\}
 =0,\\ 
 &\dot{\Pi} + \frac{2}{3}\frac{D -3}{D}p \frac{\partial v_l}{\partial x_l} + \frac{5D-6}{3D}\Pi \frac{\partial v_l}{\partial x_l} - \frac{2}{3}\frac{D-3}{D}\sigma_{\langle lk\rangle}\frac{\partial v_l}{\partial x_k}  + \frac{4(D-3)}{3D(D+2)}\frac{\partial q_l}{\partial x_l} = - \frac{1}{\tau  }\, \Pi,\\
 &\dot{\sigma}_{\langle ij\rangle} + \sigma_{\langle ij\rangle}\frac{\partial v_l}{\partial x_l} + 2\sigma_{\langle l\langle i\rangle}\frac{\partial v_{j\rangle}}{\partial x_l} - 2 (p+\Pi)\frac{\partial v_{\langle j}}{\partial x_{i\rangle}}   - \frac{4}{D+2} \frac{\partial q_{\langle i}}{\partial x_{j\rangle}} = -  \frac{1}{\tau  } \sigma_{\langle ij\rangle},\\
 & \dot{q}_i + \frac{D+4}{D+2}q_i \frac{\partial v_l}{\partial x_l} + \frac{D+4}{D+2} q_l \frac{\partial v_i}{\partial x_l} + \frac{2}{D+2}q_l \frac{\partial v_l}{\partial x_i} \\
& \qquad + \frac{D+2}{2}\frac{p}{\rho T}\left\{\left(p+\Pi + \frac{\rho}{D(D+2)p}\Delta\right)\delta_{il} - \sigma_{\langle il\rangle}\right\}\frac{\partial T}{\partial x_l}  - \frac{p}{\rho^2} \left(\Pi \delta_{il} - \sigma_{\langle il\rangle}\right)\frac{\partial \rho}{\partial x_l}  \\
&\qquad + \frac{1}{\rho}\left\{(p-\Pi)\delta_{il}+ \sigma_{\langle il\rangle} \right\}\left(\frac{\partial \Pi}{\partial x_l}  - \frac{\partial \sigma_{\langle rl\rangle}}{\partial x_r}\right) = -  \frac{1}{\tau  }\, q_i,\\
 & \dot{\Delta} + \left(\frac{D+4}{D}\Delta + 8\frac{p}{\rho}\Pi \right)\frac{\partial v_l}{\partial x_l}  - 8 \frac{p}{\rho}\sigma_{\langle ik\rangle}\frac{\partial v_i}{\partial x_k} - \frac{8}{\rho}q_i \frac{\partial p}{\partial x_i} \\
 &\qquad + 4(D+4)\frac{p}{\rho T}q_l \frac{\partial T}{\partial x_l}  
 + \frac{8p}{\rho} \frac{\partial q_l}{\partial x_l} - \frac{8}{\rho}q_i \frac{\partial \Pi}{\partial x_i} + \frac{8}{\rho} q_i \frac{\partial \sigma_{\langle il\rangle}}{\partial x_l}  =   - \frac{1}{\tau  }\Delta.
\end{split}
\label{fieldeqsMpoly}
\end{align}

By carrying out the Maxwellian iteration, from \eqref{maxwellite}, the relation between the viscosities and heat conductivity and the relaxation time is the following:
\begin{align}
 \nu = \frac{2}{3}\frac{D-3}{D}p\tau, \quad \mu = p\tau, \quad \kappa = \frac{D+2}{2}p\frac{k_B}{m}\tau,
\end{align}
and, from \eqref{deltino}, $\Delta$ is expressed as follows:
\begin{align}
 \Delta = 0.
\end{align}

 \section{Monatomic gas limit}

 The monatomic gases are described in the limit $D \rightarrow 3$ then \eqref{fieldeqsMpoly}$_4$ becomes:
\begin{align}
   \label{14pi}
&\frac{\partial \Pi}{\partial t} + v_k\frac{\partial \Pi }{\partial x_k}
    = - \left(\frac{1}{\tau  } + \frac{\partial v_k}{\partial x_k}\right) \Pi.
\end{align}
This is a first-order quasi-linear partial differential equation with respect to $\Pi$. 
As it has been studied in \cite{Arima-2013}, the initial condition for \eqref{14pi} must be compatible with the case of monatomic gas, i.e., $\Pi(0, \boldsymbol{x}) = 0$, and, assuming the uniqueness of the solution, the possible solution of Eq. \eqref{14pi} is given by
\begin{align}
   \label{Pi0}
   &\Pi(t, \boldsymbol{x}) = 0 \ \ \ (\textrm{for any} \ t).
\end{align}

If we insert the solution \eqref{Pi0} and $D=3$  into \eqref{fieldeqsMpoly}, the solutions of the present  ET$_{15}$ converge to those of  the monatomic $14$ theory given by Kremer \cite{Kremer14}.

%
%
%

 \appendix

\section{Galilean invariance of moments} \label{appendixGal}

Since the velocity independent variables are the moments in terms of the peculiar velocity $C_i=v_i - \xi_i$ instead of $\xi_i$, by inserting $\xi_i=v_i + C_i$ into \eqref{Hs}, the velocity dependence of the moments is obtained. By defining the set 
\begin{equation*}
S = \left\{ (k_1,k_2,k_3) \, | \, k_1 \geq 0, \,  k_2 \geq 0, \, k_3 \geq 0, \, k_1+k_2+k_3=s \right\},
\end{equation*}
the moments are expressed with the use of the Leibniz polynomial as follows:
\small{
\begin{align*}  
 \begin{split}
 & H^{s}_{i_1 \cdots i_h} = m  \int_{\R^{3}}
\int_0^{+\infty} f \, \left( v_{i_1} + C_{i_1} \right) \cdots \left( v_{i_h} + C_{i_h} \right)  
\,  \left(  C^2 + 2 C_i v_i + v^2 + \frac{2\mathcal{I}}{m} \right)^{s}  \dA \\
& = \sum_{p=0}^{h} \left( \begin{matrix}
h \\
p
\end{matrix}\right) 
\sum_{(k_1,k_2,k_3) \, \in S} \, \frac{s!}{k_1! \, k_2! \, k_3!} \, 
m  \inta f \, C_{( i_1} \cdots C_{i_p} v_{i_{p+1}} \cdots v_{i_h )}  
\,  \left( C^2  + \frac{2  \mathcal{I}}{m} \right)^{k_1}   \left( 2 C_i v_i \right)^{k_2} \left( v^2 \right)^{k_3}  \dA \\
&= \sum_{p=0}^{h} \left( \begin{matrix}
h \\
p
\end{matrix}\right) 
\sum_{(k_1,k_2,k_3) \, \in S} \, \frac{s!}{k_1! \, k_2! \, k_3!} 2^{k_2} \left( v^2 \right)^{k_3}  
\hat{H}^{k_1}_{j_1 \cdots j_{k_2} ( i_1 \cdots i_p} v^{ i_{p+1}} \cdots v^{i_h )}   v_{j_1} \cdots v_{j_{k_2}} \\
&= \sum_{p=0}^{h} \left( \begin{matrix}
h \\
p
\end{matrix}\right) 
\sum_{(k_1,k_2,k_3) \, \in S} \, \frac{s!}{k_1! \, k_2! \, k_3!} 2^{k_2} \left( v^2 \right)^{k_3}  
\hat{H}^{k_1}_{j_1 \cdots j_{k_2+p} }\delta_{(i_1}^{j_{k_2+1}} \cdots \delta_{i_p}^{j_{k_2+p}} v_{ i_{p+1}} \cdots v_{i_h )}   v_{j_1} \cdots v_{j_{k_2}} .  
 \end{split}
\end{align*}
}
\normalsize
We remark $0\leq k_1 \leq s$, $ p \leq h \leq N-s, k_1 + k_2 + p \leq s-k_3 + p \leq s + p \leq s + N-s = N$ and  $0\leq k_2 + p \leq s-k_1$.
By putting $k_1=r$, $k_2=k-p$ and $k_3=s+p-r-k$ and expressing the summation $\sum_{(k_1,k_2,k_3) \in S}$ as $\sum_{r=0}^{s}\sum_{k=p}^{s+p-r}$, we obtain \eqref{13}.  In fact, $0 \leq k_1 \leq s$ becomes $0 \leq r \leq s$; after that, $0 \leq k_2 \leq s-k_1$ becomes $p \leq k \leq s + p-r$. The condition $0 \leq k_3$ is automatically satisfied.


\section{Equilibrium moments} \label{appendixEq}

The equilibrium moments are
\begin{align}
 \hat{H}^{s|E}_{i_1\cdots i_h} &= m \inta f_E C_{i_1}\cdots C_{i_h} \left(C^2 + \frac{2\II}{m}\right)^s \dA \nonumber \\
&= m \sum_{q=0}^s \left(\begin{matrix} s \\ q \end{matrix}\right) \inta f^K_E f^I_E C_{i_1}\cdots C_{i_h} (C^2)^{s-q}\left(\frac{2\mathcal{I}}{m}\right)^q \dA  \nonumber \\
&= \sum_{q=0}^s  \left(\begin{matrix} s \\ q \end{matrix}\right) \left\{m\int_{\R^3}f^K_E  C_{i_1}\cdots C_{i_h} (C^2)^{s-q} d \CC \right\} \left\{ 2^q\left(\frac{k_B T}{m}\right)^q\int_0^{+\infty}f_E^I\left(\frac{\mathcal{I}}{k_B T}\right)^q \varphi(\mathcal{I})d\mathcal{I}\right\}. 
\label{Heq0}
\end{align}
In the first parenthesis we can recognize moments with respect only to the peculiar velocity that coincide with ones with the Maxwellian distribution function $\hat{F}^{r|ME}_{i_1i_2\cdots i_h}$ defined below. In fact, recalling \eqref{fint}, we have
\begin{align*}
 \hat{F}^{r|E}_{i_1i_2\cdots i_h} &= \int m f_E C_{i_1}C_{i_2}\cdots C_{i_h} (C^2)^r\dA\\
  &= \int m f_E^K C_{i_1}C_{i_2}\cdots C_{i_h} (C^2)^r d\CC = \hat{F}^{r|ME}_{i_1i_2\cdots i_h}.
\end{align*}
From the Maxwellian distribution function \eqref{fKI}$_{1}$, we obtain the explicit expression in terms of $\rho$ and $T$ as follows:
\begin{align}
 F^{r|E}_{i_1i_2 \cdots i_h} = \frac{\rho}{h+1} (2r+h+1)!! \left(\frac{k_B T}{m}\right)^{r+\frac{h}{2}}\delta_{(i_1i_2}\cdots \delta_{i_{h-1}i_h)} .
\end{align}
For example, we have
\begin{align*}
 &\hat{F}^{0|E} = \rho , \quad \hat{F}_{ij}^{0|E} = p\delta_{ij} , \quad \hat{F}_{ijrs}^{0|E} = \frac{p^2}{\rho}\left(\delta_{ij}\delta_{rs}+\delta_{ir}\delta_{js}+\delta_{is}\delta_{jr}\right),\\
 &\hat{F}_{ijrs}^{1|E}=7\frac{p^3}{\rho^2}\left(\delta_{ij}\delta_{rs}+\delta_{ir}\delta_{js}+\delta_{is}\delta_{jr}\right),\\
 &\hat{F}^{4|E} = 945 \frac{p^4}{\rho^3}.
\end{align*}


The integral in the second parenthesis in \eqref{Heq0} is a moment of the equilibrium distribution of the internal mode with respect to $\II$ that is \eqref{barI}. Then, \eqref{Heq} is obtained. 

\bigskip
\noindent

\small{
\textbf{Acknowledgments} : 
 The work has been partially supported by JSPS KAKENHI Grant Numbers JP18K13471 (TA),   by the Italian MIUR through the PRIN2017
 project "Multiscale phenomena in Continuum Mechanics:
 singular limits, off-equilibrium and transitions" Project Number:
 2017YBKNCE (SP) and GNFM/INdAM (MCC, SP and TR).
}




\end{document}